\documentclass{aastex62}
\usepackage{natbib}
\makeatletter

\newcommand{\Rmnum}[1]{\expandafter\@slowromancap\romannumeral #1@}
\makeatother
\usepackage{multirow}

\graphicspath{{./}{figures/}}

\received{December 6, 2018}
\accepted{May 4, 2019}

\shorttitle{Jet Geometry and Rate}
\shortauthors{Mogushi et al.}

\begin{document}

\title{Jet Geometry and Rate Estimate of Coincident Gamma Ray Burst and Gravitational Wave Observations}

\email{kmogushi@go.olemiss.edu}

\author[0000-0002-0786-7307]{Kentaro Mogushi}
\affil{Department of Physics and Astronomy, The University of Mississippi,\\ University MS 38677-1848, USA}

\author{Marco Cavagli\`{a}}
\affiliation{Physics Department, Missouri University of Science and Technology, \\ 1315 N.\ Pine St., Rolla MO 65409, USA}
\affiliation{Department of Physics and Astronomy, The University of Mississippi,\\ University MS 38677-1848, USA}

\author{Karelle Siellez}
\affiliation{University of Santa Cruz, Department of Astronomy and Astrophysics, \\7487 Red Hill Rd, Santa Cruz, CA 95064}
\affiliation{Georgia Institute of Technology affiliate, Center for Relativistic Astrophysics, \\837 State Street, Atlanta GA 30332}
\nocollaboration

\begin{abstract}

Short Gamma-Ray Burst (SGRB) progenitors have long been thought to be coalescing binary systems of two Neutron Stars (NSNS) or a Neutron Star and a Black Hole (NSBH). The August 17$^{\rm th}$, 2017 detection of the GW170817 gravitational-wave signal by Advanced LIGO and Advanced Virgo in coincidence with the electromagnetic observation of the SGRB GRB 170817A confirmed this scenario and provided new physical information on the nature of these astronomical events. We use SGRB observations by the Neil Gehrels Swift Observatory Burst Alert Telescope and GW170817/GRB 170817A observational data to estimate the detection rate of coincident gravitational-wave and electromagnetic observations by a gravitational-wave detector network and constrain the physical parameters of the SGRB jet structure. We estimate the rate of gravitational-wave detections coincident with SGRB electromagnetic detections by the Fermi Gamma-ray Burst Monitor to be between $\sim$ 0.1 and $\sim$ 0.6 yr$^{-1}$ in the third LIGO-Virgo observing run and between $\sim$ 0.3 and $\sim$ 1.8 yr$^{-1}$ for the LIGO-Virgo-KAGRA network at design sensitivity. Assuming a structured model with a uniform ultra-relativistic jet surrounded by a region with power-law decay emission, we find the jet half-opening angle and the power-law decay exponent to be $\theta_c\sim 7\,{}^\circ$ -- $22\,{}^\circ$ and $s\sim 5$ -- 30 at 1$\sigma$ confidence level, respectively.

\end{abstract}

\keywords{gamma-ray burst: general --- 
gravitational waves --- stars: jets --- stars: neutron --- stars: black holes}


\section{Introduction} \label{Introduction}
Gamma-Ray Bursts (GRBs) are extremely energetic electromagnetic (EM) events of astrophysical origin with prompt emission observed in the gamma-ray band. They are usually followed by an afterglow with energy ranging from the GeV to the radio band (\citealt{Meszaros:2012hj}; \citealt{Guelbenzu:2012id}). Observations show the existence of at least two classes of GRBs with distinct progenitors \citep{Kouveliotou:1993yx}. Long GRBs (LGRBs) are characterized by a softer gamma-ray emission lasting typically over two seconds. Short GRBs (SGRBs) are characterized by a harder, shorter-lived emission. While LGRB progenitors are known to be core-collapse supernovae (\citealt{Campana:2006qe}; \citealt{Lee:2004xi};    \citealt{Hjorth:2003jt}; \citealt{Fruchter:2006py}; \citealt{Woosley:2006fn}), the origin of SGRBs was long thought to be the coalescence of binary systems of two Neutron Stars (NSNS) or a Neutron Star and a Black Hole (NSBH) mergers. The recent detection of the August 17$^{\rm th}$, 2018 Gravitational-Wave (GW) signal called GW170817 by Advanced LIGO and Advanced Virgo in coincidence with the EM observation of the SGRB GRB 170817A (\citealt{TheLIGOScientific:2017qsa}; \citealt{Goldstein:2017mmi}; \citealt{Monitor:2017mdv}) confirmed the widespread hypothesis that at least some SGRBs indeed originate from NSNS mergers.
 
The LIGO Scientific Collaboration (LSC) and the Virgo Collaboration have built low-latency analysis pipelines that can promptly identify GW transient candidates (\citealt{Privitera:2013xza}; \citealt{Nitz:2017svb}). High-energy neutrino detectors and over 80 astronomical telescopes with observational capability ranging from gamma rays to the radio band have signed memoranda of understanding for the follow-up of GW detection candidates with the LSC and Virgo. Information about sky localization of a possible GW detection is distributed to these partners within few minutes from the trigger identification \citep{Branchesi:2012zs}. In parallel, the LSC and Virgo perform GW searches triggered by EM GRB observations \citep{Mandel:2011au}. The results of the search for GW signals coincident with GRBs observations by the Fermi Gamma-ray Burst Monitor (GBM) \citep{Atwood:2009ez}, the Neil Gehrels Swift Observatory (NGSO) Burst Burst Alert Telescope (BAT) \citep{Gehrels:2004aa} and the multimission detections reported through the InterPlanetary Network (IPN) \citep{Frederiks:2013cga} during the first observing run of Advanced LIGO (September 2, 2015 to January 19, 2016) were published in \citet{Abbott:2017ylp} with no evidence of GW signals coincident with SGRBs. Results of these searches for the second Advanced LIGO-Virgo observing run are expected to be released soon. Starting with the third observation run, the LSC and the Virgo Collaboration release Open Public Alerts (OPAs) for gravitational-wave transient event candidate detections.

Over the past few years, several studies have constrained the local rate density of NSNS and NSBH mergers and estimated the number of coincident observations between GW detectors and EM observatories (\citealt{Coward:2012gn}; \citealt{Petrillo:2012ij}; \citealt{Siellez:2013hia}; \citealt{Fong:2015oha}; \citealt{Guetta:2005bb}; \citealt{Chruslinska:2017odi}; \citealt{Regimbau2015}). Estimates of the local rate density of NSNS and NSBH mergers are highly uncertain, ranging from $\sim$ 10 to a few thousand events per year per cubic gigaparsec (Gpc). \citet{Coward:2012gn} estimate a local rate density $\rho_G\sim 8-1800$ Gpc$^{-3}$ yr$^{-1}$. \citet{Petrillo:2012ij}, \citet{Siellez:2013hia}, and \citet{Fong:2015oha} find larger lower bounds with local rate densities in the range $\rho_G\sim 500-1500$  Gpc$^{-3}$ yr$^{-1}$, $\rho_G\sim 92- 1154$ Gpc$^{-3}$ yr$^{-1}$, and $\rho_G\sim 90-1850$ Gpc$^{-3}$ yr$^{-1}$, respectively. \citet{Guetta:2005bb} estimate a local rate density of $\rho_G\sim 8-30$ Gpc$^{-3}$ yr$^{-1}$. All the above studies are based on EM observational data. Population synthesis studies based on the Milky Way star formation rate predict NSNS observation rates in Advanced LIGO between 2 yr$^{-1}$ \citep{Voss:2003ep} and 6 yr$^{-1}$ \citep{deFreitasPacheco:2005ub}. Studies based on the the observations of Galactic binary pulsars lead to rate estimates between 10 yr$^{-1}$ \citep{OShaughnessy:2009szr} and 35 yr$^{-1}$ \citep{Kalogera:2006uj}. A recent investigation based on simulations of compact binary evolution predicts a NSNS local rate density of $48$ Gpc$^{-3}$ yr$^{-1}$ \citep{Chruslinska:2017odi}. Null results of NSNS and NSBH merger observations in the first Advanced LIGO observing run give NSNS and NSBH merger upper bounds of $12600$ Gpc$^{-3}$ yr$^{-1}$ and $3600$ Gpc$^{-3}$ yr$^{-1}$, respectively \citep{Abbott:2016ymx}.  The NSNS merger rate estimate from the second observation run is between 340 and 4740 Gpc$^{-3}$ yr$^{-1}$ \citet{TheLIGOScientific:2017qsa}.

The detection of the GW170817/GRB 170817A provides new means of improving the above estimates and constraining the physical properties of SGRBs. The observed luminosity of GRB 170817A is lower than the observed luminosity of all SGRBs with known redshift by at least two orders of magnitude. This discrepancy could be explained by the existence of a sub-luminous population of SGRBs \citep{Siellez:2016wrh} or by GRB 170817A being observed off-axis, i.e., at a large inclination angle. \citet{Monitor:2017mdv} consider three possible scenarios for this paradigm: The ``uniform top-hat'' model \citep{Rhoads:1999wm}, where the SGRB is described by a conical jet with uniform, relativistic emission, the ``cocoon'' shock break-out model, where a quasi-isotropic emission is due to shocked material around a relativistic jet \citep{Lazzati:2017zsj}, and a ``structured jet'' model, where a narrower ultra-relativistic jet is surrounded by a mildly relativistic sheath (\citealt{Rossi:2001pk}; \citealt{Granot:2006hf}; \citealt{Pescalli:2014qja}; \citealt{Zhang:2001qt}; \citealt{Kumar:2003yt}). 
Radio and X-ray counterpart observations provide some evidence that GW170817 may be viewed off-axis \citep{Fong:2017ekk}. \citet{Margutti:2018xqd} use post-merger optical observations by the Hubble Space Telescope, radio observations by the Very Large Array and X-ray observations by the Chandra X-ray Observatory to rule out the uniform top-hat model. The Very Long Baseline Interferometric (VLBI) detection of superluminal motion in GRB 170817A also supports this conclusion \citep{Mooley:2018dlz}. The recent study for 220-260 days post-merger rules out the cocoon model in favor of a structured jet model \citep{Alexander:2018dcl}.\citet{Jin:2017hle} estimate that the number of coincident GW-EM observations for a Gaussian-type structured jet \citep{Zhang:2003uk} increase by a factor of $\sim 16$ w.r.t.\ a uniform top-hat model.

In this paper, we first use a catalog of SGRB observations by NGSO-BAT \citep{Gehrels:2004aa} with known redshifts to estimate the local rate density of NSNS and NSBH coalescences. We consider two different luminosity function models for the SGRBs, the Schechter luminosity function \citep{Andreon:2006ui} and the broken power luminosity function \citep{Guetta:2005bb}, as well as a number of different \text{star formation rate} functions (\citealt{Porciani:2000ag}, \citealt{Hernquist:2002rg}, \citealt{Fardal:2006sd}, \citealt{Cole:2000ea}, \citealt{Hopkins:2006bw}, \citealt{Wilkins:2008be}). 
We then use the observational properties of GW170817/GRB 170817A to constrain the parameters of the structured jet model \citep{Pescalli:2014qja}. Finally, we estimate the rates of GW events observable by a network of ground-based gravitational-wave detectors and the rates of coincident GW-SGRB observations with EM partners.

~

\section{Method} \label{Method}
The number $\cal N$ of SGRBs with known redshift that are {\em observed} by an EM instrument per unit observation time $t$, redshift $z$, and absolute bolometric source-frame luminosity $L$ at an inclination angle $\theta_o$ (the angle between the axis of the SGRB jet and the observer's line of sight) is
\begin{equation}
\label{eq:NGRBdiff}
{\cal N}(t,z,L,\cos\theta_o)\equiv\frac{dN(t,z,L,\cos \theta_o)}{dtdzdLd(\cos \theta_o)}=\frac{f}{1+z}\frac{dV(z)}{dz}\frac{dN_S(t',z,L,\cos \theta_o)}{dt'dVdLd(\cos \theta_o)}\,,
\end{equation}
where $N_S$ is the {\em actual} number of SGRBs with absolute luminosity $L$, inclination angle $\theta_o$ and redshift $z$ per unit comoving volume $V,$ $t'$ is the time in the SGRB local frame, and $f=f_rf_{\rm FOV}$, where $f_r$ is the fraction of observed SGRB with known redshift and $f_{\rm FOV}$ is the detector field of view. In writing Eq.\ (\ref{eq:NGRBdiff}) we have assumed SGRBs are isotropically distributed and tacitly assumed axial symmetry around the SGRB axis. If the luminosity distribution of the SGRBs is independent from the SGRB formation rate, Eq.\ (\ref{eq:NGRBdiff}) can be rewritten as
\begin{equation}
\label{eq:NGRBdiff2}
{\cal N}=\rho_s\frac{f}{1+z}\frac{dV(z)}{dz}R_S(t',z)\frac{dN_{S'}(z,L,\cos \theta_o)}{dLd(\cos \theta_o)}\,,
\end{equation}
where $R_S$ is the SGRB Rate Function (RF), i.e., the number of SGRBs per unit source time and comoving volume, and $\rho_S$ is a proportionality constant. We assume that $R_S$ is independent of $t'$ and normalize $R_S$ and $N_{S'}$ as
\begin{equation}
\label{eq:NGRBdiff3}
R_S(0)=1\,,\qquad\int_{-1}^{1}d(\cos \theta_o)\int_0^{\infty}dL\left.\frac{dN_{S'}}{dLd(\cos \theta_o)}\right|_{z=0}=1\,.
\end{equation}
With these normalizations, the constant $\rho_S$ in Eq. (\ref{eq:NGRBdiff2}) is the local rate density, i.e., the number of SGRBs per unit volume per unit time in the local universe: 
\begin{equation}
\label{eq:rhoSGRB}
\rho_S=\left.\frac{1}{f}\frac{dN}{dtdV}\right|_{z=0}\,.
\end{equation}
The number of SGRBs with known redshift up to $z$ that are observed by a given EM detector during the observation time $T_o$ is
\begin{equation}
\label{eq:cumnum}
N(z)=T_of\rho_S\int_0^z dz'\,\frac{1}{1+z'}R_S(z')\frac{dV(z')}{dz'}\int_{-1}^{1}d(\cos \theta_o)\int_{L_m (z',\cos \theta_o)}^\infty dL\,\frac{dN_{S'}(z',L,\cos \theta_o)}{dLd(\cos \theta_o)}\,,
\end{equation}
where $L_m(z,\cos \theta_o)$ is the minimum detectable luminosity of a SGRB with inclination angle $\theta_o$ and redshift $z$. The local rate density of SGRBs in Eq.\ (\ref{eq:rhoSGRB}) can be estimated by comparing the predicted theoretical value of $N(z)$ in Eq.\ (\ref{eq:cumnum}) to observations.

Throughout this paper we consider a standard flat, vacuum-dominated cosmology \citep{Spergel:2006hy}. The expression for the comoving shell in Eq.\ (\ref{eq:cumnum}) takes the form
\begin{equation}
\label{eq:co-moving}
\frac{dV(z)}{dz} = 4\pi\left(\frac{c}{H_0}\right)^3 I(z)^2\,\frac{dI(z)}{dz}\,,
\end{equation}
where
\begin{equation}
\label{eq:I_z}
I(z) = \int_0^z\frac{dz'}{\sqrt{\Omega_M (1+z')^3+\Omega_\Lambda}}\,,
\end{equation}
$c=299,792.458\, {\rm km}\, {\rm s}^{-1}$ is the speed of light in vacuum, $H_0 = 67.8(9)\, {\rm km}\, {\rm s}^{-1} \, {\rm Mpc}^{-1}$ is the Hubble constant, and $\Omega_M = 0.308\pm 0.012$ and $\Omega_\Lambda = 0.692\pm 0.012$ are the present ratio of matter and dark energy density in the Universe relative to the critical density, respectively \citep{Olive:2016xmw}. Uncertainties in the above parameters affect our final results by less than 1\% and can be safely neglected.  

Since we consider an axially symmetric structured jet emission, the SGRB absolute luminosity is related to the SGRB luminosity distance
\begin{equation}
\label{eq:luminodist}
 d_L(z)= \left(1+z\right)\frac{c}{H_0}\int_0^z dz'\,[\Omega_M \left(1+z'\right)^3+\Omega_\Lambda]^{-1/2}\,,
\end{equation}
and to the SGRB isotropic equivalent luminosity $L_I$, i.e., the luminosity that the SGRB would have if it emitted isotropically as in the direction of the observer by
\begin{equation}
\label{eq:ablumino}
L = L_{I}\int_0^1d\left(\cos\theta\right)\frac{l(\theta)}{l(\theta_o)}\,,\qquad
L_{I} = 4\pi d^2_Lk(z)F_o\,,
\end{equation}
where $l(\theta)$ is the luminosity profile, $F_o$ is the measured time-averaged energy flux in the detector's energy band and $k(z)$ is the cosmological $k$-correction factor. Under the assumption that the SGRB spectral shape is independent from the inclination angle, $F_o$ can be expressed as
\begin{equation}
    \label{eq:obsflux}
    F_o = \int^{e_2}_{e_1}Ef_{o}(E)dE=\frac{1}{\left(1+z\right)^2}\int^{e_2(1+z)}_{e_1(1+z)}Ef_s(E)dE\,,
\end{equation}
where $e_1$ and $e_2$ denote the lower and upper cutoff values of the detector's observational energy range, and $f_{o}$ and $f_s$ are the photon flux density in the observer and source frame, respectively. The cosmological $k$-correction accounts for the unobserved fraction of the source spectrum. It is given by \citep{Bloom:2001ts}
\begin{equation}
    \label{eq:k-corr}
    k = \int^{E_2}_{E_1}Ef_s(E)dE\,\bigg/\int^{e_2(1+z)}_{e_1(1+z)}Ef_s(E)dE\,,
\end{equation}
where $E_1$ and $E_2$ are the lower and upper energy values of the SGRB spectrum. We use typical values $E_1=1$ keV and $E_2=10$ MeV and the phenomenological ``Band Function'' \citep{1993ApJ...413..281B} with typical values of low- and high-energy indices $\alpha=-1$ and $\beta=-2.5$ for the source-frame photon flux density \citep{Lien:2016zny}. We use the source-frame peak energy $E^{s}_{peak}=800$ keV as suggested by \citet{Wanderman:2014eza}. With this choice, the relative differences of the time-averaged energy flux and photon flux calculated with the Power-law function and the Band function for the SGRB sample are on average $\sim$ 23\% and $\sim$ 11\%, respectively, when the photon flux density of the two functions are normalized at 50 keV \citep{Lien:2016zny}.

The number of SGRBs in Eq.\ (\ref{eq:cumnum}) depends on the metallicity of the SGRB progenitor, which is a function of the redshift (\citealt{Belczynski:2010tb}; \citealt{Belczynski:2011qp}). However, for small values of $z$ the uncertainty due to this effect is expected to be sub-dominant w.r.t.\ uncertainties arising from other factors, such as the RF. Therefore we will safely neglect the $z$ dependence in $N_{S'}$. In addition, we assume the number of SGRBs to be uniformly distributed in $\cos \theta_o$. With these assumptions, we can define the Luminosity Function (LF) $\Phi(L)$ as
%
\begin{equation}
\left.\frac{dN_{S'}}{dL}\right|_{z=0}=\Phi(L)\cos \theta_o\,.
\end{equation}
A SGRB can be detected if $F_o\geq F_m$, where $F_m$ is the minimum EM flux that can be measured in the detector's energy band. We consider the NGSO-BAT fiducial 5$\sigma$ energy flux threshold $F_m = 2.8\times 10^{-8}$ ${\rm erg\ s^{-1} cm^{-2}}$ with a duration of 1 s \citep{swifttech}. 96\% of NGSO-BAT SGRBs have the time-averaged flux above this threshold, where the time-averaged energy flux is calculated with the best-fit spectrum model \citep{Lien:2016zny}. 

The detector introduces a bias in the determination of the LF. As the instrument has a minimum detection threshold, the larger the SGRB distance the fewer low-luminosity SGRBs are observed w.r.t.\ actual distribution. Thus the fit against observational data underestimates the number of fainter SGRBs. The observed LF is obtained by rescaling the LF by the volume where the detector is sensitive \citep{Petrillo:2012ij}:
\begin{equation}
    \label{eq:rescaledlumino}
    \Phi_o(L)=\left(d_\star/d_M\right)^{-3}\Phi\,,
\end{equation}
where $d_\star$ is an arbitrary distance scale and 
\begin{equation}
\label{eq:maxdist}
d_M \propto \sqrt{L}
\end{equation}
is the maximum distance at which a SGRB of luminosity $L$ can be observed by the detector, where we have marginalized on the redshift and inclination angle. The luminosity profile of a structured jet profile with uniform emission in a cone of aperture $2\theta_c$ and power-law decay at larger angles is \citep{Pescalli:2014qja}
\begin{equation}
\label{eq:beam}
\l(\theta)=\displaystyle\left\{
  \begin{array}{@{}ll@{}}
    \displaystyle ~~1 & \text{for}\ \quad  \cos\theta_c \leq \left| \cos\theta\right| \leq 1\,, \\~\\
    \left(\displaystyle\frac{\theta\,}{\theta_c}\right)^{-s} & \text{for} \quad ~~0 \leq \left| \cos\theta\right| < \cos\theta_c\,,
  \end{array}\right.
\end{equation}
where $s>0$ and $\theta_c$ are constant parameters and $s$, $\theta_c$ are identical for all SGRBs. Throughout the paper we will refer to the emission in the region $\cos\theta_c \leq \left|\cos\theta \right| \leq 1$ as on-axis emission. As the LF is not known, we consider two different phenomenological functions (\citealt{Andreon:2006ui}, \citealt{Guetta:2005bb}). The Schechter LF is 
\begin{equation}
\label{eq:schechter} 
\Phi_o(L)= \Phi_\star\left(\frac{L\,}{L_0}\right)^{-\alpha}\,e^{-L/L_0}\qquad {\rm for}\quad  L\ge\frac{L_0}{\Delta}\,, 
\end{equation}
where $\Phi_\star$, $L_0$, $\alpha$ and $\Delta$ are constant positive parameters. $\Delta$ determines the low-luminosity cutoff of the LF. The broken power LF is 
\begin{equation}
\label{eq:BP}
\Phi_o(L)=\displaystyle\Phi_\star\left\{
  \begin{array}{@{}ll@{}}
    \left(\displaystyle\frac{L\,}{L_0}\right)^{-\alpha} & \text{for}\ \quad  \frac{L_0}{\Delta_1}\le L<L_0\,, \\~\\
    \left(\displaystyle\frac{L\,}{L_0}\right)^{-\beta} & \text{for} \quad L_0\le L<\Delta_2 L_0\,,
  \end{array}\right.
\end{equation}
where $\Phi_\star$, $L_0$, $\alpha$, $\beta$, $\Delta_1$ and $\Delta_2$ are constant positive parameters. $\Delta_1$ and $\Delta_2$ define the low- and high-luminosity cutoff of the LF, respectively. If all SGRBs in a given sample have the same inclination angle, for example they are all seen on-axis, $L$ can be replaced with $L_I$ in Eqs. (\ref{eq:schechter}) and (\ref{eq:BP}).

\begin{figure}[htbp]
 \begin{minipage}{0.5\hsize}
  \begin{center}
   \includegraphics[width=90mm]{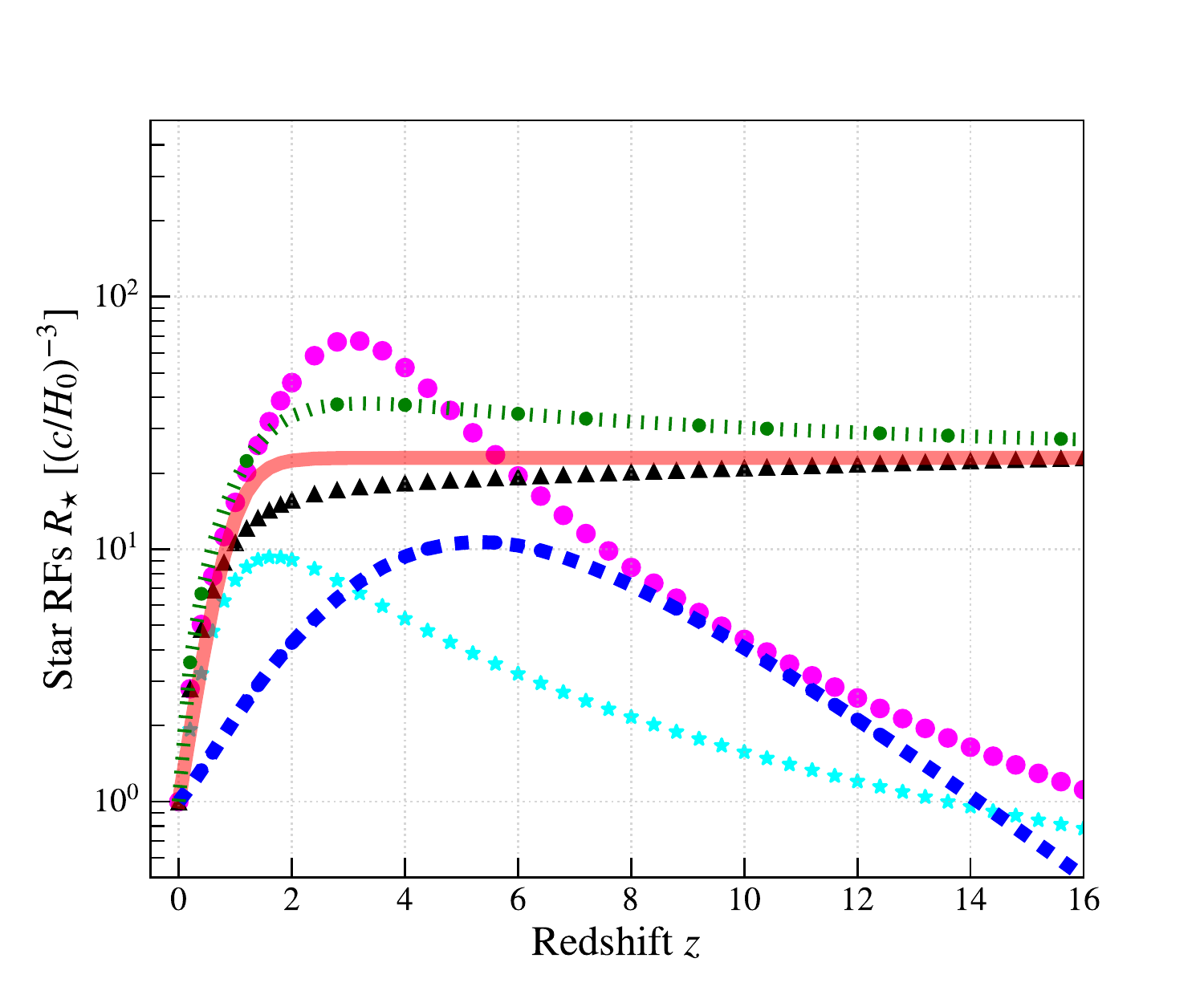}
  \end{center}
 \end{minipage}
  \begin{minipage}{0.5\hsize}
  \begin{center}
   \includegraphics[width=90mm]{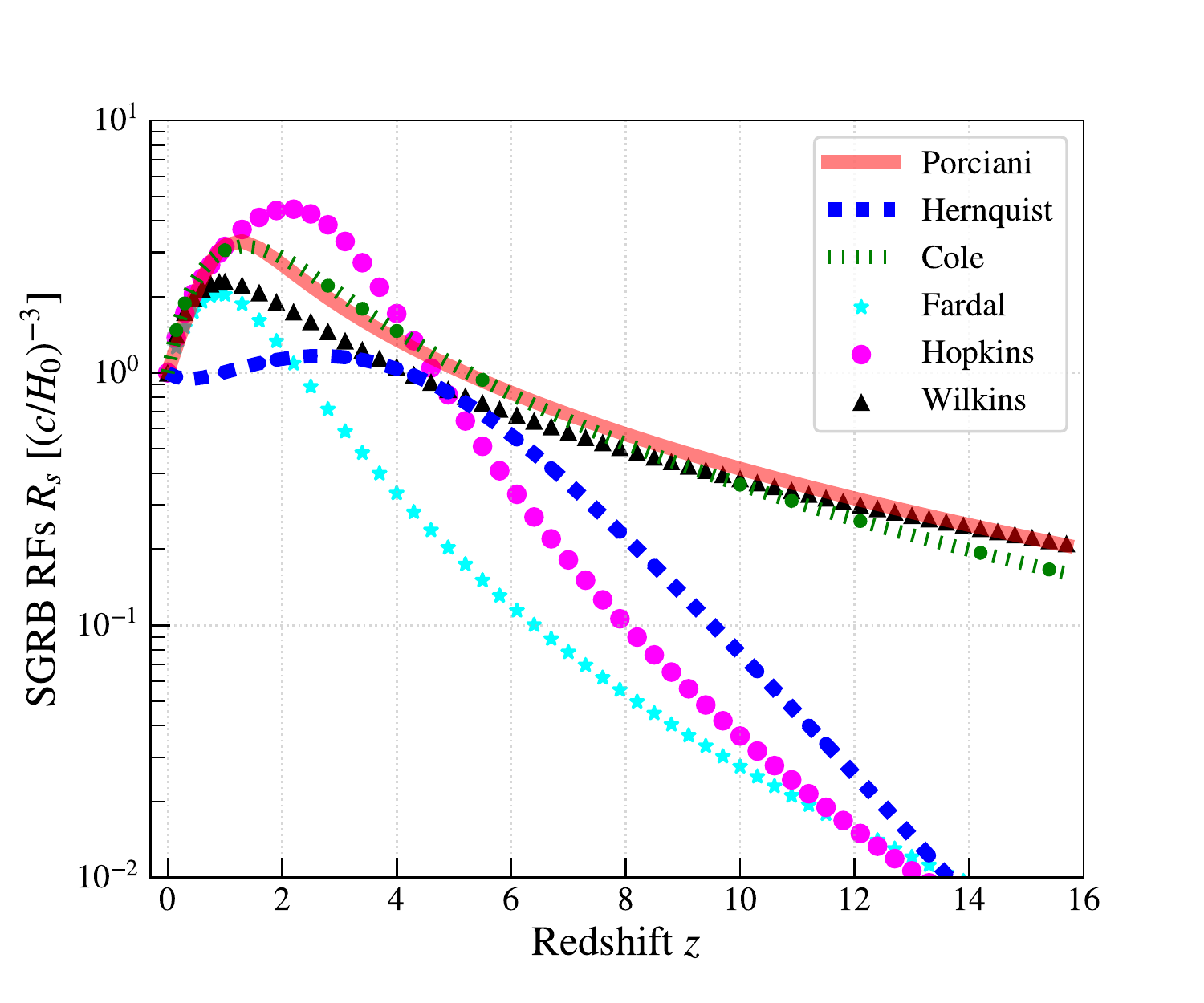}
  \end{center}
 \end{minipage}
\caption{Star RFs $R_\star$ (left panel) and SGRB RFs $R_S$ with minimum delay time $t_m=20$ Myr (right panel) in units of $(c/H_0)^{-3}$. The solid, dashed, dotted, square-marked, circle-marked, triangle-marked curves denote the Porciani, Hernquist, Cole, Fardal, Hopkins and Wilkins RFs, respectively. The RFs in the left panel are normalized to $R_{\star}(0)=1$ for the sake of comparison. The RFs in the right panel are normalized to $R_S(0)=1$. 
\label{fig:RateFunction}}
\end{figure}

The SGRB RF is expected to follow the star RF, $R_\star(z)$. However, the delay between the time of star formation and the time of the binary system coalescence affects the form of the RF. This delay time depends, among other factors, on the initial separation of the stars and the orbital eccentricity of the binary system. Therefore, the SGRB RF is given by the convolution of the star RF with the distribution of the delay time, $P(t)$ \citep{Wanderman:2014eza}. Observations of binary neutron star systems \citep{Champion:2004hc} indicate that $P(t)$ is proportional to $1/t$ with $t\gtrsim 20$ Myr \citep{Guetta:2005bb}. Studies with {\em StarTrack} population synthesis software \citep{Dominik:2012kk} suggest a delay of $\sim$ 20 (100) Myr for NSNS (NSBH) mergers. \citet{Champion:2004hc} consider a typical delay time of the order of 1 Gyr. The retarded SGRB RF is
\begin{equation}
\label{eq:retartedrate}
R_S(z) = \int_{t_m}^{T(\infty)-T(z)} dt\,\frac{1 }{1+z_\star} R_\star(z_\star)P(t)\,, 
\end{equation}
where the factor $(1+z_\star)^{-1}$ accounts for the difference between the star formation time and the coalescence time, $t_m$ is the minimum delay time, and $z_\star = Z[T(z) + t]$ is the redshift when the progenitors form. The {\em look-back time} $T(z)$ and its inverse $Z(t)$ are \citep{Hogg:1999ad} 
\begin{eqnarray}
\label{eq:lookbacktime}
T(z)&=&\frac{1}{H_0} \int_0^z\frac{dz'}{(1+z')\sqrt{\Omega_M(1+z')^3+\Omega_\Lambda}}\,,\\
Z(t)&=&\left(\frac{\Omega_\Lambda}{\Omega_M}\right)^{1/3}\left[\left(\frac{1+E(t)}{1-E(t)}\right)^2-1\right]^{1/3}-1\,,
\end{eqnarray}
where
\begin{equation}
E(t)=\exp\left[\ln\left(\frac{1+\sqrt{\Omega_ \Lambda}}{1-\sqrt {\Omega_\Lambda}}\right)-3H_0\sqrt{\Omega_\Lambda}t \right]\,.
\end{equation}
The star RF can be estimated through semi-analytical or numerical simulation methods. Both approaches require a number of assumptions on dust obscuration corrections and the stellar initial mass function \citep{Wilkins:2008be}. As a result, different models may predict quite different RFs. In the following, we define $R_\star =\tilde R_\star(c/H_0)^{-3}$ and consider six different star formation models:
\begin{enumerate}
\item \textbf{CHW} (\citealt{Cole:2000ea}, \citealt{Hopkins:2006bw}, \citealt{Wilkins:2008be}):
\begin{equation}
\label{eq:colehopwil}
\tilde R_\star(z)=R_0\frac {a+bz}{1+\left(z/c\right)^d} H(z)\,,\qquad H(z)=H_0\sqrt{(1+z)^3\Omega_{\rm M }+ \Omega_{\rm \Lambda}} \,,
\end{equation}
where $a=0.0166$, $b=0.1848$, $c=1.9474$, $d=2.6316$ for the Cole model, $a=0.0170$, $b=0.13$, $c=3.3$, $d=5.3$ for the Hopkins model, and $a=0.014$, $b=0.11$, $c=1.4$, $d=2.2$ for the Wilkins model, respectively. 

\item \textbf{Fardal} \citep{Fardal:2006sd}:
\begin{equation}
\label{eq:fardal}
\tilde R_\star(z)=R_0\frac{(1+z)^{p_2}}{\left[1+p_{\rm 1}(1+z)^{p_2}\right]^{p_3+1}}H(z)\,,
\end{equation}
where $p_1=0.075$, $p_2=3.7$, and $p_3=0.84$.

\item \textbf{Porciani} \citep{Porciani:2000ag}:
\begin{equation}
\label{eq:Porciani}
\tilde R_\star(z)=R_0\frac{e^{3.4z}}{e^{3.4z}+22}\,.
\end{equation}

\item \textbf{Hernquist} \citep{Hernquist:2002rg}:
\begin{equation} \label{eq:hernquist}
\tilde R_\star(z)=R_0\frac {\chi^2}{1+\alpha(\chi-1)^3e^{\beta\chi^{7/4}}}\,,\qquad \chi (z)=\left[\frac{H(z)}{H_0 }\right]^{2/3}\,,
\end{equation}
where $\alpha = 0.012$, $\beta=0.041$.

\end{enumerate}
The normalization constants $R_0$ in the previous equations are chosen so that $R_s(0)=1$. The RFs are shown in Fig.\ \ref{fig:RateFunction}. 

\section{Luminosity function and jet geometry} \label{constrain}

\begin{figure}[htbp]
    \begin{minipage}{0.5\hsize}
  \begin{center}
   \includegraphics[width=90mm]{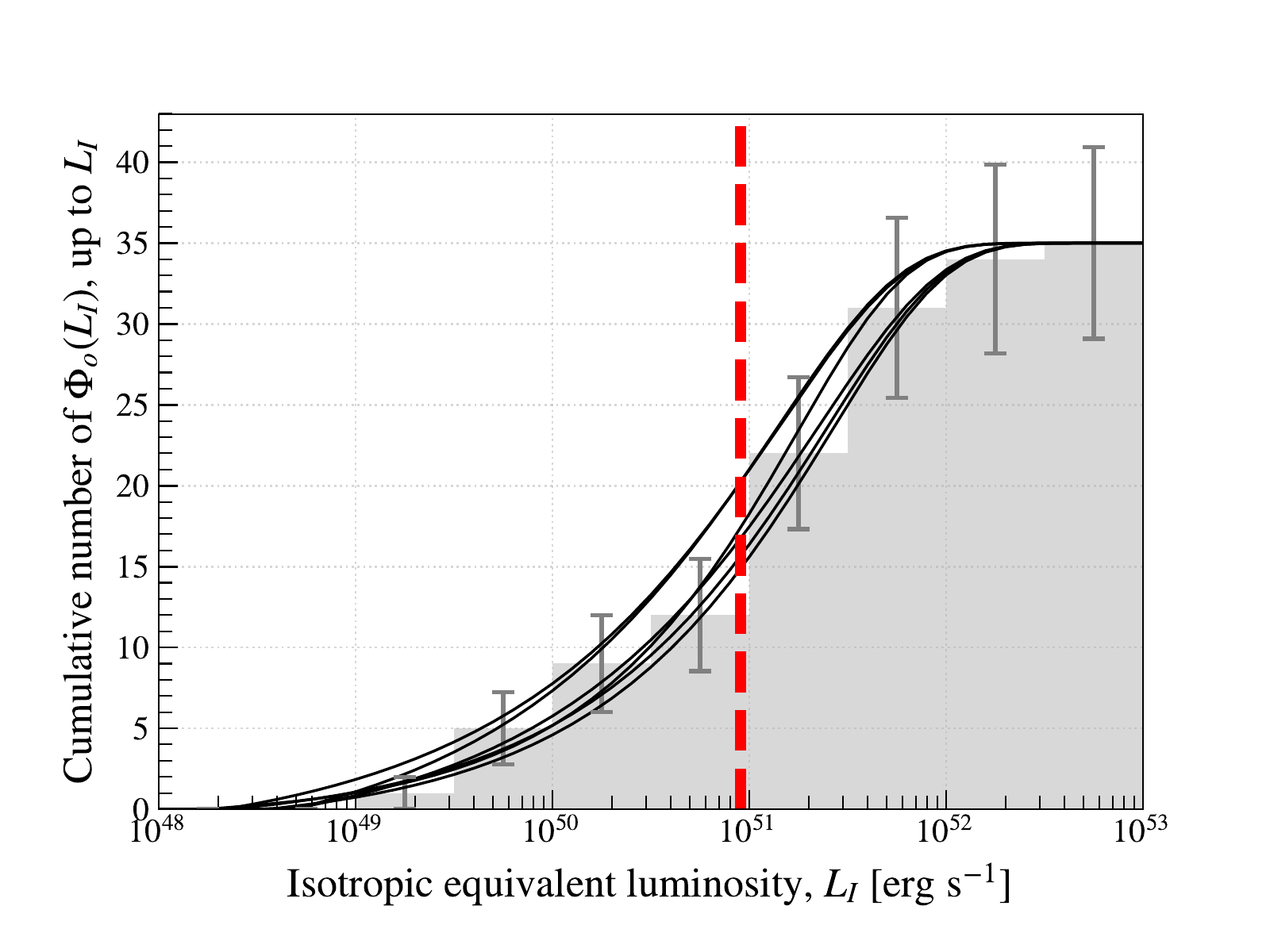}
  \end{center}
    \end{minipage}
    \begin{minipage}{0.5\hsize}
  \begin{center}
   \includegraphics[width=90mm]{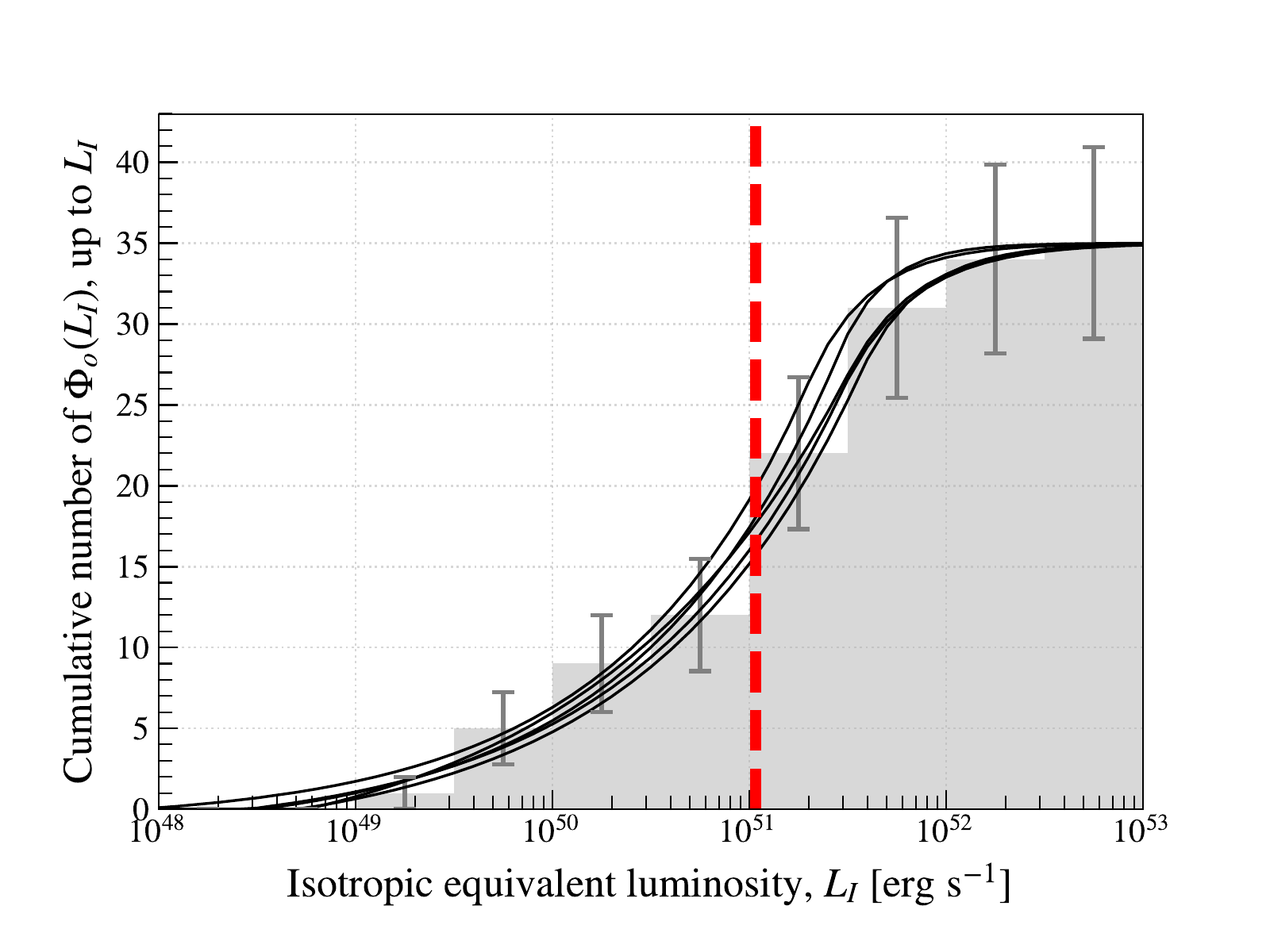}
  \end{center}
    \end{minipage}
\caption{Best fits for the Schechter (left) and broken power (right) cumulative LFs, $\Phi_o(L_I)$. The curves denote least-square best fits with widths passing a chi-square two-tailed test at 90\% confidence level for different widths of the histogram bins in log scale, $\Delta l\equiv\log_{10}(L_n/L_{n-1})$, where $\Delta l \in \left[0.1, 1.0\right]$ and $n$ is the bin index. Error bars in the histograms denote statistical uncertainties. The red dashed vertical lines give the average of the medians calculated for each LF best fit.}
\label{fig:luminosity}
\end{figure}

Appendix B lists the SGRBs used in our analysis. Following \citet{Fong:2017ekk}, we assume that all SGRBs in the sample were observed on-axis. We evaluate the parameters of the LF with the exception of $\Delta_2$ in Eq. (\ref{eq:BP}) by fitting Eqs.\ (\ref{eq:schechter}) and (\ref{eq:BP}) against the cumulative number of SGRBs in our sample. The value of $\Delta_2$ does not significantly affect the determination of the other parameters provided that $\Delta_2\gg 1$. \citet{Guetta:2005bb} choose $\Delta_2 = 10^2$. For sake of computational efficiency, we set $\Delta_2 = 10^3$. The best fits of the LF models are shown in Fig.\ \ref{fig:luminosity} for different choices of bin widths. The best-fit parameters are summarized in Table \ref{table:paralumino}.
\begin{table}[h!]
\begin{center}
\begin{tabular}{c c c c c } 
\hline
	&	$p_1$	&	$p_2$	&	$p_3$	&	$p_4$\\
\hline\hline
Schechter	&	51.6$\pm$0.2 	&	0.55$\pm$0.07	&	--	&	3.2$\pm$0.2	\\
Broken power &	51.5$\pm$0.1	&	0.60$\pm$0.05	&	2.4$\pm$0.3	&	3.0$\pm$0.3\\
\hline
\end{tabular}
\caption{Best-fit parameters for the Schechter and broken power LFs. The parameters are ($p_1$, $p_2$, $p_4$)=($\log_{10}L_0$, $\alpha$, $\log_{10}\Delta$) for the Schechter LF and ($p_1$, $p_2$, $p_3$, $p_4$)=($\log_{10}L_0$, $\alpha$, $\beta$, $\log_{10}\Delta_1$) for the broken power LF. The values in the table are obtained by averaging on the parameters of each curve shown in Fig.\ \ref{fig:luminosity}. Uncertainties are standard deviations.}
\label{table:paralumino}
\end{center}
\end{table}

\begin{figure}[htbp]
  \begin{center}
   \includegraphics[width=100mm]{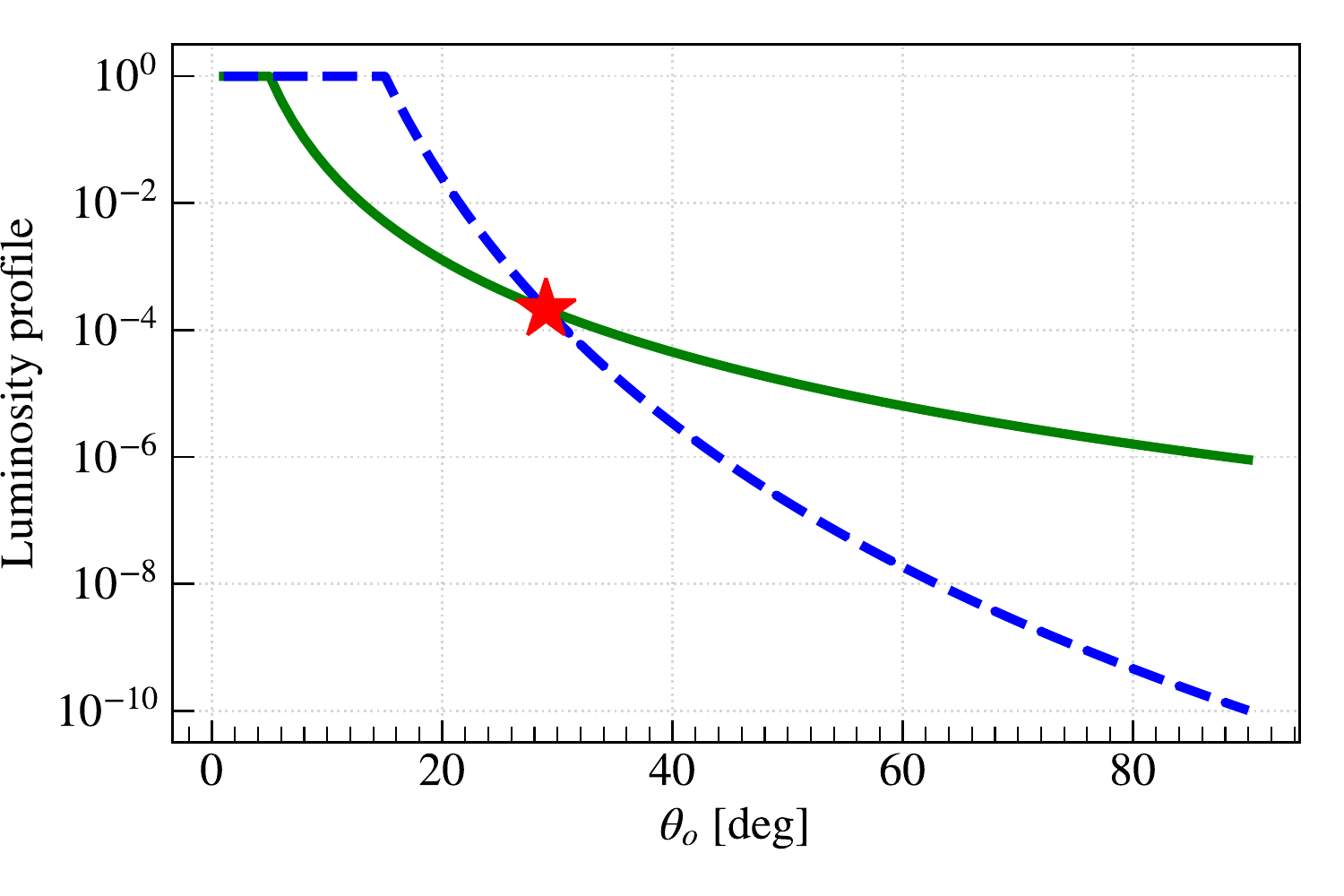}
   \end{center}
\caption{Luminosity profile of GRB 170817A as a function of the inclination angle for two choices of the jet parameters. The solid green and dashed blue curves give the luminosity profile for $(\theta_c, s)=(5\,{}^{\circ},4.8)$ and $(\theta_c, s)=(15\,{}^{\circ},12.8)$, respectively. The star denotes the isotropic equivalent luminosity of GRB 170817A normalized to the typical SGRB isotropic equivalent luminosity for an inclination angle of $\theta_G=29^\circ$, corresponding to the median value of the probability density from \citet{Abbott:2018wiz}.}
\label{fig:relativeflux}
\end{figure}

Using the LF and GW170817/GRB 170817A observational data, we can constrain the SGRB flux parameters $\theta_c$ and $s$. Since GRB 170817A is seen off-axis, its (measured) isotropic equivalent luminosity must be rescaled to compare it to the isotropic equivalent luminosity of the SGRBs in the NGSO-BAT sample (which are seen on-axis). Using Eq.\ (\ref{eq:beam}) in Eq.\ (\ref{eq:ablumino}) the absolute luminosity of a SGRB in the NGSO-BAT sample can be written as 
\begin{equation}
\label{eq:L_GRB}
L = L_I\eta(\theta_c,s)\,,\qquad\eta = 1-\cos\theta_c+\int_0^{\cos\theta_c}d(\cos\theta)\left(\frac{\theta}{\theta_c}\right)^{-s}\,.
\end{equation}
The absolute luminosity $\cal L$ of GRB 170817A can be written as
\begin{equation}
\label{eq:L_GW}
    {\cal L} = {\cal L}_I\eta\left(\frac{\theta_G}{\theta_c}\right)^{s}\,,
\end{equation}
where ${\cal L}_I$ and $\theta_G$ are the isotropic equivalent luminosity and inclination angle of GRB 170817A, respectively. If we assume that GRB 170817A is a typical SGRB, we can substitute ${\cal L}=\langle L_I\rangle\eta$ in Eq.\ (\ref{eq:L_GW}), where $\langle L_I\rangle$ is the typical SGRB luminosity. Solving for $s$, we find
\begin{equation}
\label{eq:s}
s = \frac{\ln\left[\langle L_I\rangle/{\cal L}_{I}\right]}{\ln \left[{\theta_G}/\theta_c\right]}\,.
\end{equation}
Equation (\ref{eq:s}) allows us to set constraints on the $(\theta_c,s)$ parameter space. We estimate ${\cal L}_{I}=2.3\times 10^{47}$ erg s$^{-1}$ by converting the time-average flux observed by Fermi-GBM without soft-tail emission, ${\cal F}_{\rm GBM}\sim 3.1\times 10^{-7}$ erg s$^{-1}$ cm$^{-2}$, to the flux in NGSO-BAT's spectrum range, ${\cal F}_{\rm NGSO}\sim 1.5\times 10^{-7}$ erg s$^{-1}$ cm$^{-2}$ and using GW observational data for the inclination angle and the luminosity distance $d_{\cal L}=40$ Mpc \citep{TheLIGOScientific:2017qsa}. The estimate of the inclination angle depends on several assumptions, most notably the spins of the neutron stars. Very Long Baseline Interferometric (VLBI) observations suggest an inclination angle of $20\,{}^\circ \pm 5\,{}^\circ $ \citep{Mooley:2018dlz}. In the following we will conservatively consider a larger range of values $\theta_G = 15\,{}^\circ - 40\,{}^\circ$, which is consistent with the 90\% c.l. interval given in \citet{Abbott:2018wiz}. The typical SGRB isotropic equivalent luminosity $\langle L_{I}\rangle$ is estimated by calculating the median of each LF best fit in Fig.\ (\ref{fig:luminosity}) and then averaging over the values. Figure \ref{fig:relativeflux} shows two examples of flux profiles compatible with Eq. (\ref{eq:s}).  

The assumption that GRB 170817A is seen off axis implies $\theta_c < \theta_G$. The LIGO network observed one NSNS merger in the second observation run (O2). The probability densities $P_c$ and $P_s$ for $\theta_c$ and $s$  can be expressed in terms of the truncated Poisson distribution 
\begin{equation}
P_{\{c,s\}}(x) = \displaystyle\frac{\lambda(x) e^{-\lambda(x)}\left|\displaystyle\frac{d\lambda}{dx}\right|}{\displaystyle\int_0^a dx\  \lambda(x) e^{-\lambda(x)}\left|\displaystyle\frac{d\lambda}{d x}\right|}\,,
\label{eq:dist_theta_c}
\end{equation}
where $a=\theta_G$ $(a=\infty)$ for $x=\theta_c$ $(x=s)$ and $\lambda = \mathcal{R}\rho_G\mathcal{V}_{LH}T_G$ is the expected number of NSNS mergers detected in the O2 search volume $\mathcal{V}_{LH}$ and observing time $T_G$, 
\begin{equation}
    \rho_G = \rho_S/\left[\Gamma_{NSNS} \mathcal{R} + \Gamma_{NSBH}(1-\mathcal{R})\right]   
\end{equation}
is the local rate density of the sum of NSNS and NSBH mergers, where $\mathcal{R}$ is the fraction of NSNS mergers to the total number of NSNS+NSBH mergers, $\Gamma_{NSNS}$ and $\Gamma_{NSBH}$ are the fraction of NSNS and NSBH mergers that produce SGRBs, respectively. The effective range for NSNS mergers in O2 was $\mathcal{V}_{LH}= 88$ Mpc with an effective observation time of $T_G \sim$ 0.3 years.

\citet{TheLIGOScientific:2017qsa} estimate the rate of local NSNS mergers to be between 340 and 4740 Gpc$^{-3}$ yr$^{-1}$. The NSBH rate is highly uncertain, but null detection in the LIGO first observation run (O1) gives an upper bound of $\sim$ 3600 Gpc$^{-3}$ yr$^{-1}$ \citep{Abbott:2016ymx}. We consider a range for $\mathcal{R}$ from $\mathcal{R}=0.5$, corresponding to a NSBH merger rate compatible with the NSNS merger rate, to $\mathcal{R}= 1$, corresponding to a NSBH merger rate equal to zero. We set $\Gamma_{NSNS}=1$ and consider $\Gamma_{NSBH}$ in the the range 0.1-0.3 \citep{Stone:2012tr}. We estimate the local rate density $\rho_G$ for each different LF, RF, $t_m$, and jet parameters and then average over the LF best fits. We use NGSO's observation time of 12.6 years with a duty cycle factor of 78\%, corresponding to $T_o=9.8$ years \citep{Lien:2016zny} and averaged BAT's field of view of 1.4 sr \citep{Barthelmy:2005hs}, corresponding to $f_{\rm FOV}=0.1$. 107 SGRBs were observed during 12.6 years, leading to $f_r=35/107$ \citep{Lien:2016zny}. As the various RFs are comparable for $z \lesssim 1$ (see Fig.\ \ref{fig:RateFunction}), the choice of RF does not significantly affect the overall result. Therefore, for the sake of illustration, we only present the results for the Hernquist and Hopkins RFs with $t_m=$ 100 Myr and 20 Myr, respectively.

Larger values of $\mathcal{R}$ lead to a larger expected number of NSNS mergers in O2. Greater values of $\rho_G$ imply a smaller half aperture angle to match observed SGRB population. For instance, the value of $\theta_c$ calculated with $\mathcal{R}=1$ is 10\% greater than the value calculated with $\mathcal{R}=0.5$. Similarly, larger values of $\Gamma_{NSBH}$ imply a lower number of NSNS mergers in O2. A value of $\theta_c$ obtained with $\Gamma_{NSBH}=0.1$ is 3\% larger than the value obtained with $\Gamma_{NSBH}=0.3$.

In the following, we choose as representative values $\Gamma_{NSBH}=0.2$ and $\mathcal{R}=5/6$, the latter corresponding to the median value of the local rate density of NSNS mergers from \citet{TheLIGOScientific:2017qsa} and the median value of the local rate density of NSBH mergers from \citet{Abbott:2016ymx}.

\begin{figure}[htbp]
    \begin{minipage}{0.5\hsize}
  \begin{center}
   \includegraphics[width=100mm]{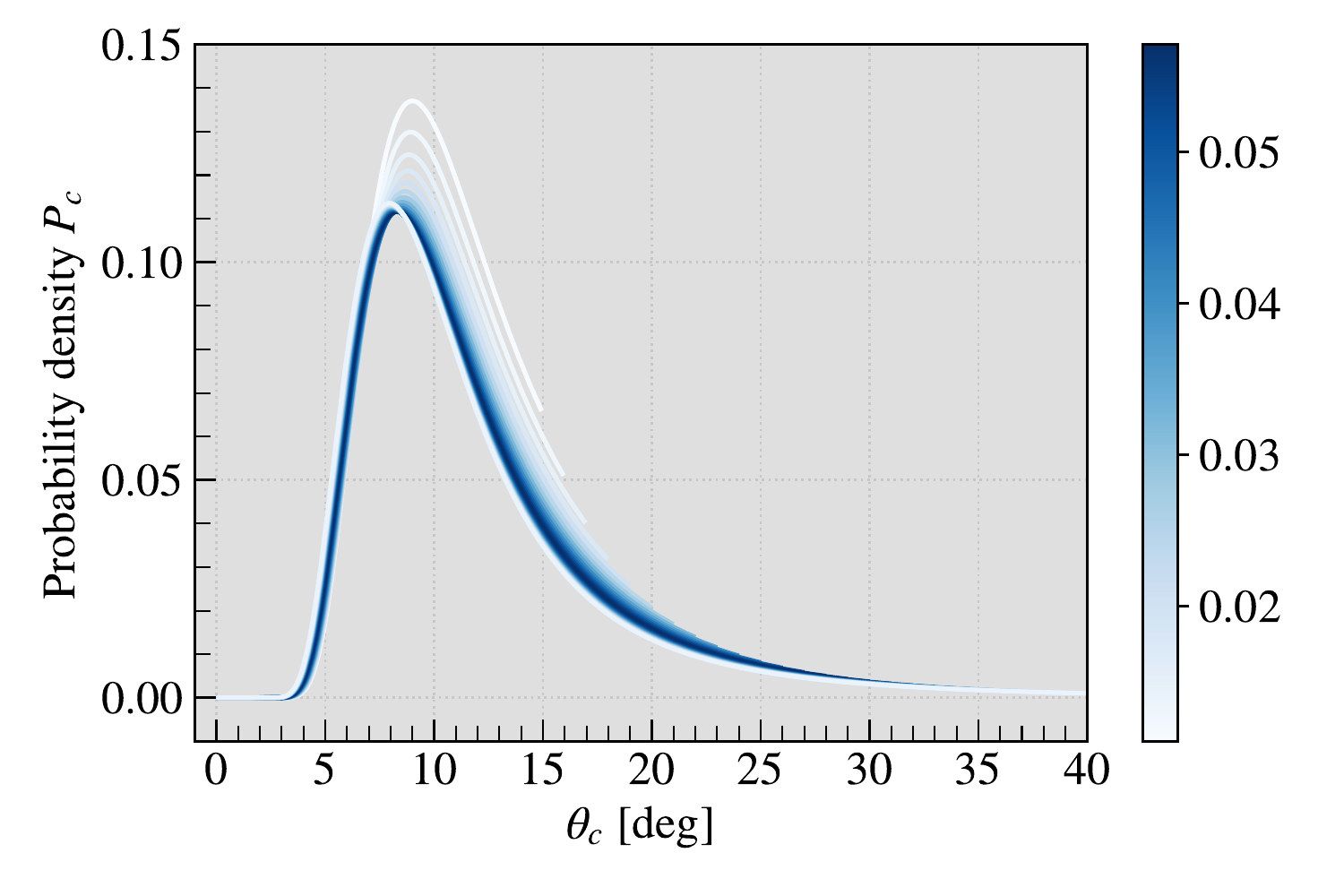}
  \end{center}
    \end{minipage}
    \begin{minipage}{0.5\hsize}
  \begin{center}
   \includegraphics[width=100mm]{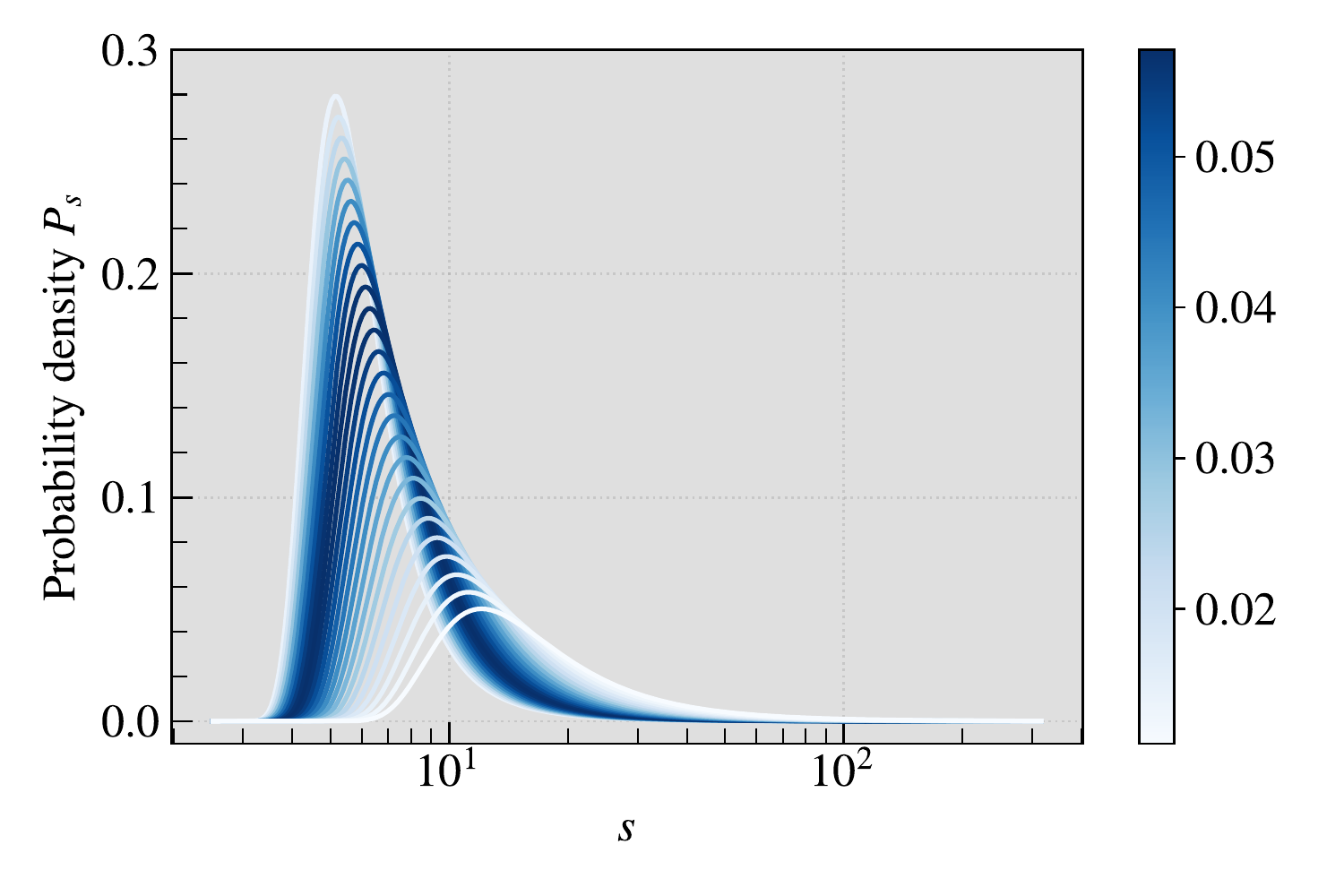}
  \end{center}
    \end{minipage}
\caption{Probability densities of the jet parameters $P_c$ (left panel) and $P_s$ (right panel) for the broken power LF, the Hopkins RF and delay time $t_m$ = 20 Myr and different values of $\theta_G\in [15\,{}^{\circ}, 40\,{}^{\circ}]$. The probability densities are obtained by averaging over the LF best fits of Fig. \ref{fig:luminosity}. The color scale indicates the probability density of the observed GRB 170817A inclination angle from \citet{Abbott:2018wiz}.}
\label{fig:chance_beampara}
\end{figure}

Figure \ref{fig:chance_beampara} shows the jet parameter probability densities for different values of the GRB 170817A inclination angle. The color scale denotes the probability density of the GRB 170817A inclination angle for the ``PhenomPNRT'' waveform model with low-spin prior (see Fig.\ 4 in \citealt{Abbott:2018wiz}). The most likely values of $\theta_c$ are comparable throughout any values of inclination angle $\theta_G$. However, the values of $s$ are smaller for larger values of $\theta_G$. Hence, a larger $\theta_G$ increases the chance of detecting off-axis SGRBs relative to on-axis SGRBs. 
 
\begin{figure}[htbp]
  \begin{center}
   \includegraphics[width=120mm]{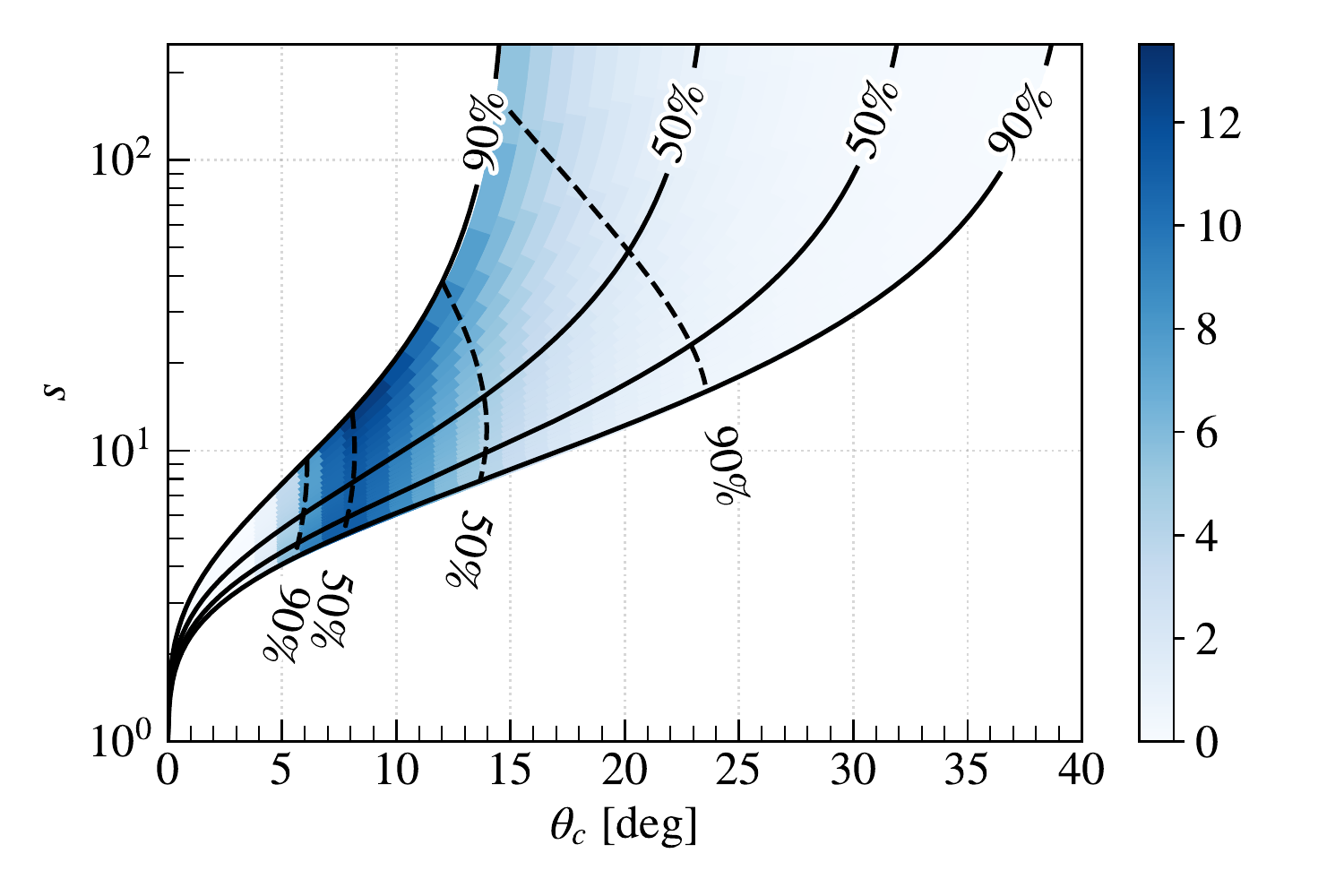}
  \end{center}
\caption{Allowed region of the jet parameter space. The colored region bounded by the solid black curves represent the allowed regions of the jet parameter space obtained by considering the 90\% and 50\% c.l.\ intervals for the observed GRB 170817A inclination angle \citep{Abbott:2018wiz}. The values of $\theta_G$ increase from left to right. The color scale represents the probability for the opening angle $\theta_c$. The dashed black curves give the lower and upper bounds at 90\% and 50\% c.l.}
\label{fig:regions}
\end{figure}

Figure \ref{fig:regions} shows the allowed region of the ($\theta_c,s$) parameter space. The colored region bounded by the solid and dashed blue curves represent the allowed region for the 90\% and 50\% c.l.\ intervals of the observed GRB 170817A inclination angle, respectively. The color scale represents the probability of having a given opening angle $\theta_c$ with the solid and dashed cyan curves denoting 90\% and 50\% c.l.\ intervals. Table.\ \ref{table:theta_s} shows values of $\theta_c$ and $s$ for different LFs, RFs, $t_m$ and $\theta_G=29\,{}^\circ$, that is the median value of the probability density of inclination angle in \citet{Abbott:2018wiz}. 

\begin{table}[htbp!]
\renewcommand*{\arraystretch}{1.2}
\begin{center}
\begin{tabular}{c c c c c } 

\hline
	LF		&	RF	&	$t_{m}$ [Myr]	&	$\theta_c$ [ $^\circ$]	&	$s$\\
\hline\hline
\multirow{4}{*}{Schechter}		&	\multirow{2}{*}{Hernquist} 	&	100	    & $13.0_{-4.0}^{+6.8}$	&	$10.3_{-3.2}^{+11.5}$			\\
								&								&	20		& $12.9_{-4.0}^{+6.8}$	&	$10.3_{-3.2}^{+11.3}$		\\
                                \cline{2-5}
                               	&	\multirow{2}{*}{Hopkins}	&	100		& $9.5_{-3.0}^{+5.6}$	&	$7.4_{-1.9}^{+5.3}$		\\
                                &                               &	20		& $9.0_{-2.8}^{+5.3}$	&	$7.1_{-1.7}^{+4.7}$			\\
\hline                                
\multirow{4}{*}{Broken power}	&	\multirow{2}{*}{Hernquist}	&	100		& $14.8_{-4.5}^{+6.9}$	&	$12.6_{-4.4}^{+16.8}$			\\
								&								&	20		& $14.8_{-4.5}^{+6.9}$	&	$12.5_{-4.3}^{+16.6}$			\\
                                \cline{2-5}
								&	\multirow{2}{*}{Hopkins}	&	100		& $11.0_{-3.4}^{+6.2}$	&	$8.7_{-2.4}^{+7.4}$			\\
								&								&	20		& $10.3_{-3.2}^{+5.9}$	&	$8.2_{-2.2}^{+6.5}$			\\
\hline            
\end{tabular}
\caption{Best estimates of the jet parameters for different LFs, RFs, and minimum delay times. The values in the table are the median values of the $P_c$ and $P_s$ distributions calculated with $\theta_G=29\,^\circ$. Quoted uncertainties are at 68\% c.l.}\label{table:theta_s}
\end{center}
\end{table}
\newpage
\section{Local rate density and number of coincident events} \label{localandcoincident}

Using the values of $\theta_c$ from Table \ref{table:theta_s} we can estimate the local rate density of GW events and the projected number of observations by a network of GW detectors. The local rate density varies between $\rho_G=1100\pm 1000$ Gpc$ ^{-3}$ yr$^{-1}$ and $\rho_G=4500\pm 4300$ Gpc$ ^{-3}$ yr$^{-1}$, where the lower (upper) value is obtained from the 1$\sigma$ larger (smaller) value of $\theta_c$ in Table \ref{table:theta_s} and the uncertainties follow from the $1\sigma$ uncertainties from averaging over the LF best fits. The median value of $\theta_c$ for the various models gives $\rho_G=2400$ Gpc$ ^{-3}$ yr$^{-1}$, which can be considered as the best estimate for the local rate density. The uncertainty in the local rate density is mainly due to the uncertainty in the determination of the jet opening angle. Smaller values of $\theta_c$ imply fewer observable SGRBs and a larger number of actual binary system coalescences to match observations. For instance, $\theta_c = 7\,{}^{\circ}$ for the model with the Hopkins RF, the broken power LF and 20 Myr delay time leads to a local rate density which is $\sim$ 4 times larger than the local rate density calculated with $\theta_c = 16\,{}^{\circ}$.

To see how the various RFs, LFs, and delay times affect the local rate density estimate, we arbitrarily fix the jet opening angle to $\theta_c=10\,{}^{\circ}$ and vary all other parameters. The Hopkins RF is characterized by a higher SGRB formation rate at small $z$ w.r.t.\ other RFs (see right panel of Fig.\ \ref{fig:RateFunction}), thus implying a lower local rate density to fit observations. The Hernquist RF typically gives local rate densities about 1.8 times larger than the local rate densities obtained with the Hopkins RF (see right panel of Fig.\ \ref{fig:RateFunction}). Shorter minimum delay times imply smaller initial orbital separations of the compact objects and a faster evolution of the binary system towards coalescence. As the number of SGRBs tends to peak at larger $z$, shorter minimum delay times lead to smaller local rate densities. A minimum delay time $t_m=20$ Myr gives a local rate density approximately 90\% smaller than the local rate density obtained with $t_m=100$ Myr. The broken power LF leads to local rate densities 29\% larger than the local rate densities obtained with the Schechter LF. The broken power LF predicts a larger population of intrinsically faint SGRBs than the prediction of the Schechter LF, suggesting the existence of a larger population of faint distant SGRBs that may escape detection. For example, the minimum cutoff luminosity of the broken power LF obtained by averaging the LF best fits in Fig. \ref{fig:luminosity} is $\sim$ 1\% smaller than the minimum cutoff luminosity of the Schechter LF.

\begin{figure}[htbp]
  \begin{center}
   \includegraphics[width=100mm]{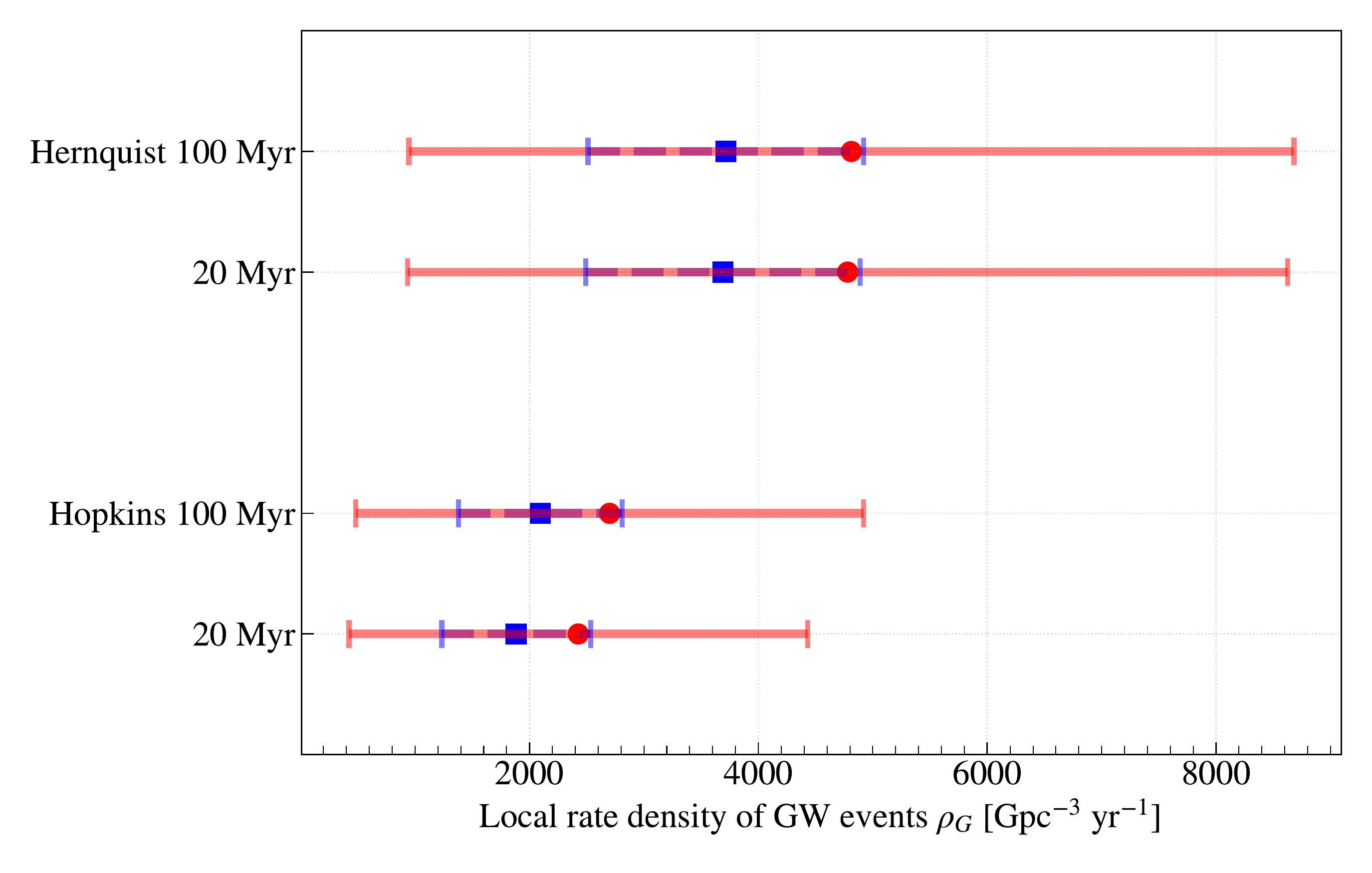}
  \end{center}
\caption{Local rate density calculated for different RFs, minimum delay times and LFs (red = broken power, blue = Schechter) with a fixed jet opening angle $\theta_c=10\,{}^{\circ}$ and $\theta_G=29\,^\circ$. The red circles and blue squares denote the average values of $\rho_G$ for the broken power and the Schechter LFs, respectively. The bars designate 1$\sigma$ uncertainties due to the LFs. The Hernquist RF and the Schechter LF give local rate densities typically higher than the Hopkins RF and the broken power LF.}
\label{fig:LRD}
\end{figure}

Given the local rate density, we can estimate the number of GW events and the number of coincident GW-SGRB events observable by a network of GW detectors and EM partners. Let us define the duty cycle factor of a network comprising ${\cal N}\geq2$ GW detectors as
\begin{equation}
\label{eq:duty}
\mathcal{D}_{i_1\cdots i_{\cal N}} \equiv \prod_{k=1}^{\cal N}\left[D_k^{i_k}\left(1-D_k\right)^{1-i_k} \right]\,,
\end{equation}
where $k=1,\cdots,{\cal N}$ is a label uniquely identifying the detectors, $D_k$ is the duty cycle factor of the $k^{\rm th}$ detector, and $i_k=(0\,,1)$ indicates whether the $k^{\rm th}$ detector is in observing mode ($i_k=1$) or not observing mode ($i_k=0$). (See Appendix B for a derivation of Eq.\ (\ref{eq:duty}) and following equations.) For example, in the case of the Advanced LIGO network comprising the LIGO-Livingston (LLO) detector with duty cycle factor $D_1$ and the LIGO-Hanford (LHO) detector with duty cycle factor $D_2$, the network duty cycle factor is 
\begin{equation}
\mathcal{D}_{i_{1}i_{2}}=\left\{
  \begin{array}{@{}ll@{}}
    D_1D_2 & \quad\hbox{LLO observing and LHO not observing}\,, \\
    D_1(1-D_2) & \quad\hbox{LLO observing and LHO not observing}\,,\\
    (1-D_1)D_2 & \quad\hbox{LLO not observing and LHO observing}\,,\\
    (1-D_1)(1-D_2) &\quad\hbox{LLO and LHO not observing}\,.\\
  \end{array}\right.
\end{equation} 
The fraction of time a given subset of $m$ detectors are in observing mode is given by Eq.\ (\ref{eq:duty}) with $i_k=1$ for $k\in\{m\}$ and $i_k=0$ for $k\notin\{m\}$. Using Eq.\ (\ref{eq:duty}), the total number of NSNS and NSBH mergers that can be simultaneously observed by at least two detectors in the network is
\begin{equation}
\label{eq:numberLIGO}
N_G = \rho_GT_G \sum\limits_{i_1=0}^1\cdots\sum\limits_{i_{\cal N}=0}^1 \mathcal{D}_{i_1\cdots i_{\cal N}} \left[ \mathcal{R} \mathcal{V}_{i_1\cdots i_{\cal N}}+(1-\mathcal{R})\mathcal{U}_{i_1\cdots i_{\cal N}}\right]\,,
\end{equation}
where $T_G$ is the network running time and $\mathcal{V}_{i_1 \cdots i_{\cal N}}$ ($\mathcal{U}_{i_1 \cdots i_{\cal N}}$) is the second largest single-detector NSNS (NSBH) search volume when at least two detectors are in observing mode (zero otherwise). Similarly, the number of mergers that are observable by at least two GW detectors in coincidence with an EM detector is 
\begin{equation}
\label{eq:coin}
N_C = \frac{T_G D_{\rm EM} f_{\rm EM}}{fT_o}\sum\limits_{i_1=0}^1\cdots\sum\limits_{i_{\cal N}=0}^1 \mathcal{D}_{i_1\cdots i_{\cal N}} \left[\mathcal{P}N\left(\mathcal{Z}_{i_1\cdots i_{\cal N}}\right) + \left(1-\mathcal{P}\right)N\left(\mathcal{Y}_{i_1\cdots i_{\cal N}}\right)\right]\,,
\end{equation}
where
\begin{equation}
    \mathcal{P} = \frac{\Gamma_{NSNS}\mathcal{R}}{\Gamma_{NSNS}\mathcal{R} + \Gamma_{NSBH}(1-\mathcal{R})}
\end{equation}
is the fraction of SGRBs that are produced by NSNS mergers, $D_{\rm EM}$ is the duty cycle of the EM detector, $f_{\rm EM}$ is its field of view, and $N(\mathcal{Z}_{i_1\cdots i_{\cal N}})$ and $N(\mathcal{Y}_{i_1\cdots i_{\cal N}})$ are the numbers of SGRB from Eq.\ (\ref{eq:cumnum}) up to redshifts $\mathcal{Z}_{i_1\cdots i_{\cal N}}$ and $\mathcal{Y}_{i_1\cdots i_{\cal N}}$, corresponding to the search volumes $\mathcal{V}_{i_1\cdots i_{\cal N}}$ and $\mathcal{U}_{i_1\cdots i_{\cal N}}$, respectively. The total number of mergers that are observable by at least a single GW detector in coincidence with an EM detector can be obtained by setting $\mathcal{Z}_{i_1\dots i_{\mathcal{N}}}$ and $\mathcal{Y}_{i_1\dots i_{\mathcal{N}}}$ to the largest single-detector NSNS and NSBH detection redshifts, respectively.

To estimate the number of coincident events detectable between Fermi-GBM and a GW detector we set the field of view for Fermi-GBM to 70\%, the duty cycle factor to 85\% \citep{Burns:2015fol}. Following \citet{Burns:2015fol}, we treat Fermi-GBM and NGSO-BAT as equally sensitive. To calculate the time-averaged energy flux threshold $1.0 \times 10^{-7}$ erg cm$^{-2}$ s$^{-1}$ for Fermi-GBM in the energy band 10--1000 keV, we convert the fiducial energy flux threshold for NGSO-BAT from the observer-frame Band function, which is obtained using the source-frame Band function with the mean value $z=0.69$ from \citet{Wanderman:2014eza}. About 94\% of GBM SGRBs have the time-averaged flux above this threshold.

We assume conservative NSNS inspiral ranges of 120 Mpc for the LIGO detectors and 65 Mpc for Virgo in the third observing run (O3), and 190 Mpc for LIGO, 65 Mpc for Virgo, and 40 Mpc for KAGRA in the fourth observing run (O4) \citep{Aasi:2013wya}. We also consider the scenario with NSNS inspiral ranges of 190 Mpc for LIGO, 125 Mpc for Virgo, and 140 Mpc for KAGRA at design sensitivity. We set the duty cycle factor of each detector to 80\%. We assume the inspiral range for NSBH mergers to be approximately 1.6 times larger than the inspiral range of NSNS mergers \citep{Aasi:2013wya}. The predicted rates of combined NSNS and NSBH mergers observable by at least two GW detectors in O3, O4 and at design sensitivity, and the corresponding predicted rates of coincident events observable by NGSO-BAT and Fermi-GBM are summarized in Table \ref{table:BNSandNSBHtable}.

\begin{table}[h!]
\begin{center}
\begin{tabular}{c c c c c} 
\hline 
\multicolumn{2}{ c }{Observing run [Network]}	 &  O3[LHV] & O4[LHV] & Design[LHKV]\\
\hline\hline
$N_G/{\rm yr}$	&two GW detectors 	&  $4-66$  & $15-251$ &	$17-296$\\
\hline
\multirow{2}{*}{$N_C/{\rm yr}$}	&  NGSO-BAT (two GW detectors) 	& $0.002-0.02$ 	&	$0.01-0.1$ & $0.05-0.2$  \\
 &	Fermi-GBM (two GW detectors)				& $0.1-0.6$ & $0.3-1.6$ & $0.3-1.8$\\
\hline
\multirow{2}{*}{$N_C/{\rm yr}$}	&  NGSO-BAT (single GW detector) 	& $0.02-0.1$ 	&	$0.06-0.3$ & $0.06-0.3$  \\
 &	Fermi-GBM (single GW detector)				& $0.1-0.8$ & $0.4-2.2$ & $0.4-2.2$\\
\hline
\end{tabular}
\caption{Estimated rates of combined NSNS and NSBH GW detections and coincident EM observations per calendar year of network's observation time by at least two GW detectors or a single GW detector in O3, O4 and at the detector design sensitivity. The results are derived for the model with the broken power LF, Hopkins RF with delay time 20 Myr, and assuming a duty cycle factor of 80\% for each GW detector. Ranges in the table are obtained by varying the jet half-opening angle in the 1$\sigma$ interval $[7.1\,{}^{\circ}, 16.2\,{}^{\circ}]$ (see Table \ref{table:theta_s}) and the broken power LF best fits (see Fig.\ \ref{fig:luminosity}). L, H, V, and K stand for LIGO-Hanford, LIGO-Livingston, Virgo, and KAGRA, respectively. Fermi-GBM and NGSO-BAT teams are looking for sub-threshold weak GRB signals that are missed with standard trigger criterion around the time when GW events are detected. Sub-threshold detection could potentially increase $N_C$.}
\label{table:BNSandNSBHtable}
\end{center}
\end{table}

\begin{figure}[htbp]
  \begin{center}   \includegraphics[width=100mm]{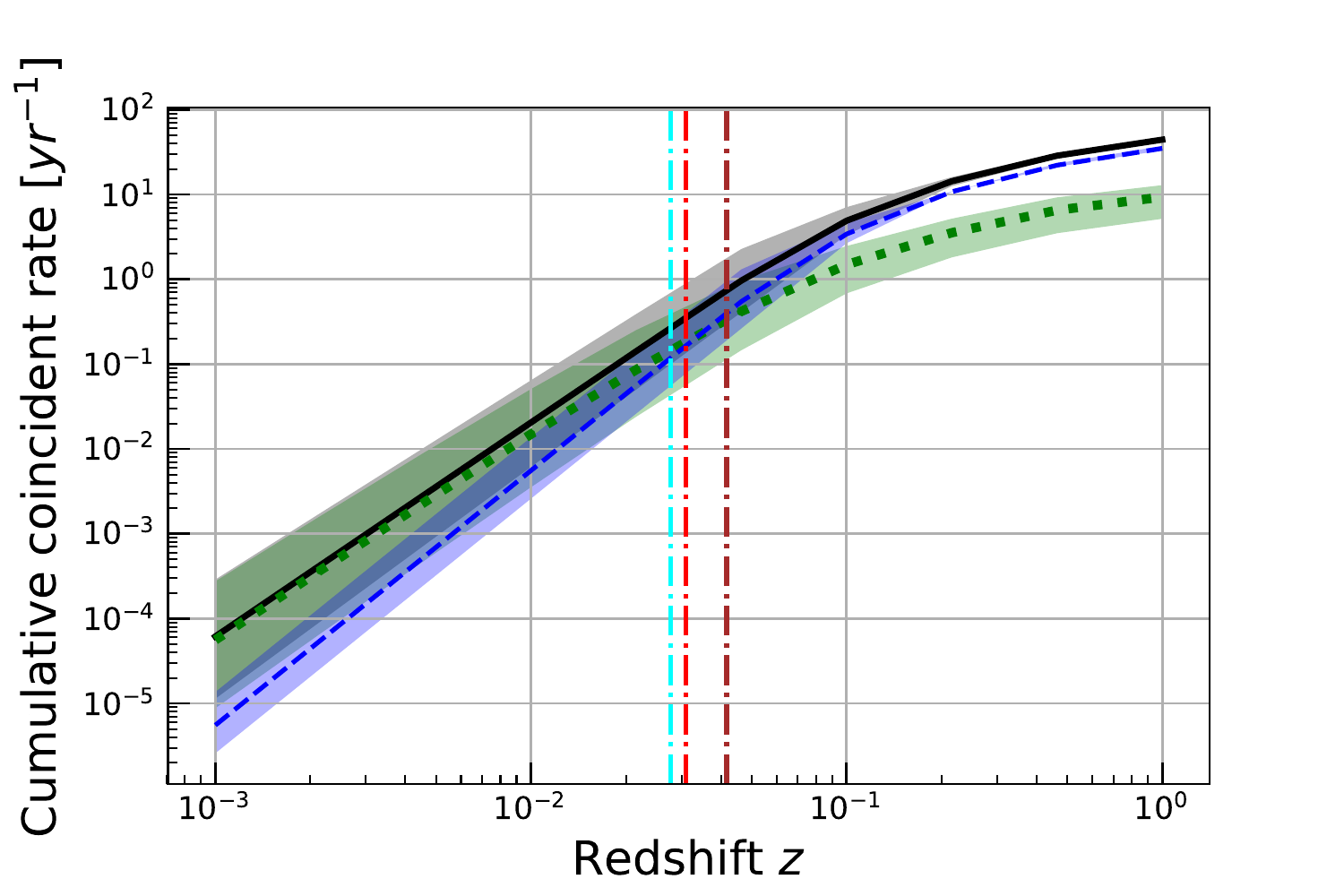}
  \end{center}
\caption{Rate of coincident events per calendar year detectable by Fermi-GBM and a single GW detector for the model with the Hopkins RF, minimum delay time 20 Myr, broken power LF, $\theta_c=10.3\,{}^\circ\,^{+5.9\,{}^{\circ}}_{-3.2\,{}^{\circ}}$ and $\theta_G=29\,{}^\circ$ as a function of the redshift. The Fermi-GBM field of view and duty cycle factor are assumed to be 85\% of the sky coverage and 70\%, respectively. The duty cycle factor for the GW detector is 80\%. The dashed-blue curve denotes the rate of on-axis events, the dotted-green curve denotes the rate of off-axis events, and the solid-black curve denotes the total rate. The dot-dashed red, cyan, and gray curves represent the NSNS inspiral ranges for LIGO, Virgo and KAGRA at design sensitivity, respectively. The shaded areas denote 1$\sigma$ uncertainties in the estimates due to $\theta_c$ and the LF best fits (see Fig. \ref{fig:luminosity}).}
\label{fig:Ccum}
\end{figure}

The rate of on-axis SGRB events can be obtained by replacing the lower bound of the $\cos \theta_o$ integral in Eq.\ (\ref{eq:cumnum}) with $\cos\theta_c$, multiplying by a factor two and choosing $f$ equal to the Fermi-GBM field of view. Figure \ref{fig:Ccum} shows the rate of on-axis, off-axis, and total coincident events per calendar year detectable by Fermi-GBM and a single generic GW detector as a function of $z$. The fraction of off-axis detections for the LHV and the LHKV networks is estimated to be between 50\% and 85\% in O3, and 45\% and 65\% in O4 and at design sensitivity. As the NSBH search volume is greater than the NSNS search volume (by approximately a factor of 4), different choices of $\mathcal{R}$ and $\Gamma_{NSBH}$ lead to different estimates on the number of coincident observations. Smaller values of $\mathcal{R}$ imply a larger population of NSBH mergers and larger values of $\Gamma_{NSBH}$ imply greater fractions of NSBH mergers producing SGRBs. For example, $\mathcal{R}=0.5$ gives $N_C$ $\sim$ 44\% larger than the value obtained with $\mathcal{R}=1$ and $\Gamma_{NSBH}=0.3$ gives $N_c\sim 9$\% larger than the value obtained with $\Gamma_{NSBH}=0.1$. The largest number of coincident events in O3, $N_C\sim 0.8$, is obtained with $\mathcal{R}=0.5$ and $\Gamma_{NSBH}=0.3$.

\section{Discussion and conclusion} \label{conclusion}

Using a catalog of SGRB observations by NGSO-BAT \citep{Gehrels:2004aa} and EM and GW data from GW170817/GRB 170817A, we have estimated the local rate density of NSNS and NSBH coalescences and derived constraints on the geometry of SGRB jets. Our data sample comprises 35 SGRBs with known redshift that were observed by NGSO-BAT in a $\sim 12$ year period, from December 17$^{\rm th}$, 2004 to June 12$^{\rm th}$, 2017. We considered the Schechter and broken power models for the LF, and various models for the RF with different delay times. 

We find that the typical value of the half-opening angle $\theta_c$ in a structured jet profile \citep{Pescalli:2014qja} is between $7\,{}^\circ$ and $22\,{}^\circ$ with the power-law decay exponent $s$ varying between 5 and 30 at 1$\sigma$ confidence level. Using these results, the local rate density of GW events across all considered models is estimated to be between $\rho_G=1100\pm 1000$ Gpc$ ^{-3}$ yr$^{-1}$ and $\rho_G=4500\pm 4300$ Gpc$ ^{-3}$ yr$^{-1}$. 

The choices of the LF and the RF affect these results. The broken power LF implies a larger population of low-luminous SGRBs. Thus models with the broken power LF lead to half-opening angles greater than those predicted by models with the Schechter LF. Narrower (wider) jet opening angles imply a larger (smaller) local rate density. For example, $\theta_c = 7\,{}^{\circ}$ leads to a local rate density $\sim$ 4 times larger than the local rate density calculated with $\theta_c = 16\,{}^{\circ}$. The choice of RF is the factor which most affects the results. Different RFs lead to different estimates because of assumptions about the initial stellar mass functions, dust obscuration corrections and minimum delay times. For example, the median value of the half-opening angle ranges from $9\,{}^\circ$ for the Hopkins RF to $15\,{}^\circ$ for the Hernquist RF. The model with the broken power LF, Hopkins RF, and the minimum delay time $t_m=20$ Myr can be used as a representative example with most likely values $\theta_c=10.3\,{}^{\circ}{}^{+5.9\,{}^{\circ}}_{-3.2\,{}^{\circ}}$ and $s=8.2^{+6.5}_{-2.2}$. The rate of GW observations in O3 (O4) Advanced LIGO-Virgo observation run for this model is between $N_G\sim4$ (17) and 66 (251) events per calendar year with a rate of coincident GW-SGRB observations by Fermi-GBM and at least two GW detectors in the network between $N_C \sim 0.1$ (0.3) and 0.6 (1.6). About $50-85$\% ($45-65$\%) of these events in O3 (O4) are expected to be detections of off-axis GRBs. The corresponding values for the LIGO-Virgo-KAGRA network at design sensitivity are between $N_G \sim 17$ and 296, and $N_C \sim 0.3$ and 1.8 with a fraction of off-axis GRBs comparable to O4. If only one GW detector is required to claim a coincident event, values increase by $\sim$ 40\% in O3 and O4 and by $\sim$ 20\% at network design sensitivity.
    
The composition of the SGRB sample may also affects the results. As a consistency check, we calculated the number of coincident GW-Fermi GBM observations in O3 without considering GRB 090417A and GRB 070923, whose localization is $\sim$ 60 times worse than the localization of the other SGRBs in the sample. We found that the results did not significantly change ($N_C \sim 0.05$ -- 0.6) as their effect on the determination of the LF is minimal.

As a second consistency check, we also estimated the number of coincident observations by the two-detector LIGO network and Fermi-GBM in the latest O2 run and found $N_C\sim$ 0.02 -- 0.1. In deriving this result, we assumed GRB 170817A to be a typical SGRB, i.e., we set the absolute luminosity of GRB 170817A equal to the median of the SGRB sample. If the actual absolute luminosity of GRB 170817A is lower, the jet profile must decay more slowly in order to match the observed luminosity. As the emission at wider angles provides the dominant contribution for detections at low $z$, the estimated upper bound of coincident GW-Fermi GBM observations could then increase. For example, by choosing the absolute luminosity of GRB 170817A one order of magnitude lower, we find $\theta_c = 9.8\,{}^{\circ}{}^{+6.1\,{}^\circ}_{-3.2\,{}^\circ}$, $s=5.7^{+4.6}_{-1.5}$ and an estimated upper bound of coincident GW-Fermi GBM observations $\sim 0.2$. Different choices of $\mathcal{R}$ and $\Gamma_{NSBH}$ can also lead to a larger $N_C$. If we set $\mathcal{R} = 0.5$ and $\Gamma_{NSBH}=0.3$, the number of coincident GW-Fermi GBM observations in O2 can be as high as $\sim 0.3$ and between 0.2 (0.5) and 1.0 (2.0) for O3 (O4). KAGRA's contribution will be negligible, as its sensitivity in O4 is expected to be low enough to affect results only by a few percent $\sim$ 1\%. The above results are in agreement with the estimate of $N_C=0.1$ -- 1.4 events per year given in \citet{Monitor:2017mdv}, where an extended power law LF with minimum isotropic luminosity of $10^{47}$ erg s$^{-1}$ is assumed to either account for off-axis dimmer events or the presence of a larger, low-luminosity population of SGRBs.

Recently, several other independent investigations have provided estimates for the expected rate of coincident GW and EM observations in future observing runs. Using Monte Carlo simulations with a structured jet model from \citet{Margutti:2018xqd}, \citet{Gupte:2018pht} estimate the percentage of coincident NSNS GW and Fermi-GBM observations to be about 30\%.  The larger number of coincident observations is mainly due to the assumption of a uniform inclination angle, which increases the chance of detecting on-axis emissions. \citet{Beniamini:2018uwo} consider several jet models, including a structured jet model, the Gaussian jet model, and a cocoon-like model to show that GRB 170817A is atypical. They estimate the number of coincident GW-Fermi GBM observations to be $\sim$ 1 per year within a distance of 220 Mpc. \citet{Bhattacharya:2018lmw} estimate the rates of coincident GW and EM detections (prompt and cocoon emission) to be in the range $0.8 - 4.4$ per year at Advanced LIGO design sensitivity for a wide range of NSBH merger parameters such as mass, spin and NS equation of state. \citet{Howell:2018nhu} perform a Bayesian inference using GRB 170817A EM data and the Gaussian jet model to predict coincident rates of $0.2-1.8$ per year in O3 and $0.3-4$ per year at design sensitivity. Both Fermi-GBM and NGSO-BAT teams are looking for sub-threshold weak GRB signals that are missed with standard trigger criterion around the time when GW events are detected, which can potentially lead to additional coincident observations. No matter what nature decides to offer us, new coincident detections in upcoming observing runs will certainly allow us to refine these estimates and better constrain the geometry of the SGRB jets.

\acknowledgments
\section*{Acknowledgments}
This work has been supported by NSF grants PHY-1921006, PHY-1707668 and PHY-1404139. The authors would like to thank colleagues at the LIGO Scientific Collaboration and the Virgo Collaboration for their help and useful comments. The authors are grateful for computational resources provided by the LIGO Laboratory and supported by NSF grants PHY-0757058 and PHY-0823459. K.M.\ would like to thank all the LIGO Laboratory staff and LSC fellows for their support during his stay at LIGO Hanford Observatory, where part of this work was completed, and colleagues at the University of Mississippi, Shrobana Ghosh, Sumeet Kulkarni and Dripta Bhattacharjee for their help.



\appendix
\section{SGRB data sample}
In our analysis, we consider the sample of SGRBs with known redshift from \citet{swiftarchive} and extend it to include SGRBs with redshift obtained through the observation of an afterglow, as shown in \citep{Siellez:2016wrh}.

\begin{table}[!h]
\begin{center}
\begin{tabular}{c || c | c | c| c } 
\hline
SGRB name & Flux [$10^{-7} \: {\rm erg} \: {\rm cm}^{-2}$] & Redshift &	$L_{I}$[erg s$^{-1}$]	& Reference  \\ [0.5ex] 
\hline\hline
161104A	&		3.66	&	0.788	&	$5.4\times10^{51}$	&	\Rmnum{1}			\\
160821B	&		2.15	&	0.16	&	$1.1\times10^{50}$&	\Rmnum{2}	\\
160624A	&		1.68	&	0.483	&   $8.8\times10^{50}$&	{\em NGSO}		\\
150423A &		2.83	&	1.394 	&	$1.4\times10^{52}$&	{\em NGSO}  \\
150120A	&		1.06	&	0.46 	&	$5.1\times10^{50}$&	{\em NGSO} \\
150101B	&		0.881	&	0.1343	&	$3.2\times10^{49}$&	\Rmnum{3} \\
141212A	&		2.16	&	0.596 	&	$1.8\times10^{51}$&	{\em NGSO} \\
140903A	&		3.75	&	0.351 	&	$1.0\times10^{51}$&	{\em NGSO} \\
140622A	&   	0.868	&	0.959	&	$1.9\times10^{51}$&	{\em NGSO}	\\
131004A	&		1.35	&	0.717 	&	$1.6\times10^{51}$&	{\em NGSO}\\
130603B	&		24.9	&	0.3565 	&	$6.9\times10^{51}$&	\Rmnum{4}\\
120804A	&		8.82	&	1.3		&	$3.7\times10^{52}$&	\Rmnum{5}\\
111117A	&		2.92	&	1.3 	&	$1.2\times10^{52}$&	\Rmnum{6},\Rmnum{7} \\
101219A	&		3.86	&	0.718 	&	$4.7\times10^{51}$&	{\em NGSO}\\
100724A	&		1.03	&	1.288	&   $4.2\times10^{51}$&	{\em NGSO}	\\
100628A	&		5.27	&	0.102	&	$1.1\times10^{50}$&	\Rmnum{8}\\
100625A	&		5.91	&	0.452 	&	$2.7\times10^{51}$&	\Rmnum{9}\\
100206A	&		10.5	&	0.407 	&	$3.9\times10^{51}$&	\Rmnum{10}\\
100117A	&		2.89	&	0.915 	&	$5.8\times10^{51}$&	\Rmnum{11}\\
090927	&		0.82	&	1.37	&   $3.8\times10^{51}$&	{\em NGSO}\\
090515	&		4.65	&	0.403 	&	$1.7\times10^{51}$&	\Rmnum{12}\\
090510	&		0.971	&	0.903 	&	$1.9\times10^{51}$&	{\em NGSO}\\
090426	&		1.35	&	2.609 	&	$2.2\times10^{52}$&	{\em NGSO}\\
090417A	&		1.75	&	0.088 	& 	$2.7\times10^{49}$&	\Rmnum{13}\\
080905A	&		1.3	    &	0.1218 	&   $3.8\times10^{49}$&	\Rmnum{14}\\	
070923 	&   	8.77	&	0.076 	&	$9.9\times10^{49}$&	\Rmnum{15}\\
070729	&	    0.9	    &	0.8		&	$1.4\times10^{51}$&	\Rmnum{16}\\
070724A &		0.633	&	0.457	&	$3.0\times10^{50}$&	{\em NGSO}\\
070429B	&		1.22	&	0.9023 	&	$2.4\times10^{51}$&	\Rmnum{17}\\
061217	&		1.67	&	0.827 	&	$2.7\times10^{51}$&	{\em NGSO}\\
061201	&		4.0	    &	0.111 	&	$9.8\times10^{49}$&	{\em NGSO}\\
060801	&		1.47	&	1.1304 	&	$4.6\times10^{51}$&	\Rmnum{18}\\
060502B &		2.8	    &	0.287	& 	$4.9\times10^{50}$&	{\em NGSO}	\\
051221A	&		5.46	&	0.547 	&	$3.7\times10^{51}$&	{\em NGSO}\\
050509B &		2.55	&	0.225 	&	$2.7\times10^{50}$&	{\em NGSO}\\
\hline
\end{tabular}
\caption{\small List of the SGRBs used in our analysis. Time-averaged energy flux is taken by \citet{Lien:2016zny}. $L_{I}$ is the isotropic equivalent luminosity. The last column gives the reference for the redshift estimate: NGSO data center archive \citep{swiftarchive} ({\em NGSO}); \cite{FongChornock:2016} (\Rmnum{1}); \cite{Levan:2016} (\Rmnum{2}); \cite{Fong:2016irn} (\Rmnum{3}); \cite{Postigo:2013rua} (\Rmnum{4}); \cite{Berger:2012ay} (\Rmnum{5}); \cite{Sakamoto:2012vm} (\Rmnum{6}); \cite{Margutti:2012aj} (\Rmnum{7}); \cite{2015A...583A..88N} (\Rmnum{8}); \cite{Fong:2013eqa} (\Rmnum{9}); \cite{Perley:2011kk} (\Rmnum{10}); \cite{Fong:2010kz} (\Rmnum{11}); \cite{Berger:2010ag} (\Rmnum{12}); \cite{2009GCN..9136....1O}, \cite{2009GCN..9134....1F}, \cite{2009GCN..9137....1B} (\Rmnum{13}); \cite{Rowlinson:2010jb} (\Rmnum{14}); \cite{FoxOfek} (\Rmnum{15}); \cite{Leibler:2010uq} (\Rmnum{16}); \cite{2008arXiv0802.0874C} (\Rmnum{17}); \cite{Berger:2006ik} (\Rmnum{18}).}
\label{table:datasample}
\end{center}
\end{table}
\newpage
\section{Network duty cycle factor}

In this appendix we derive Eqs.\ (\ref{eq:duty})-(\ref{eq:coin}). Let us define $\mathcal{R}$ as the fraction of NSNS mergers to the total number of NSNS and NSBH mergers, and $T_G$ as the running time of a network comprising  ${\cal N}\geq 2$ GW detectors. The fraction of the observation time of the $k^{\rm th}$ detector is proportional to $D_k$ or $(1-D_k)$ if the detector is observing or not, where $D_k$ is the duty cycle factor of the $k^{\rm th}$ detector and $k = 1,\dots\mathcal{N}$ is a label uniquely identifying the detectors. The duty cycle factor of the network is given by Eq.\ (\ref{eq:duty}). By summing Eq.\ (\ref{eq:duty}) on all possible combinations of $i_k=(0,1)$, where $i_k=1$ ($0$) indicates that the $k^{\rm th}$ detector is (not) in observing mode, we find
\begin{eqnarray}
\sum\limits_{i_1=0}^{1}\cdots \sum\limits_{i_{\cal N}=0}^{1}{\cal D}_{i_1\cdots i_{\cal N}} &=&  \sum\limits_{i_1=0}^{1}\cdots \sum\limits_{i_{\cal N}=0}^{1} \left\{\prod_{k=1}^{\cal N}\left[D_k^{i_k}(1-D_k)^{1-i_k} \right]\right\} \nonumber \\
&=& \prod_{k=1}^{\cal N}\left\{\sum\limits_{i_k = 0}^1 \left[ D_k^{i_k}(1-D_k)^{1-i_k}\right]\right\} \nonumber \\
&=& \prod_{k=1}^{\cal N}\left[(1-D_k) + D_k\right] \nonumber \\
& = & 1\,,
\end{eqnarray}
as expected. The fraction of time a given subset of $m$ detectors are in observing mode is given by Eq. (\ref{eq:duty}) with $i_k=1$ for $k \in {m}$ and $i_k=0$ for $k \notin {m}$:
\begin{equation}
T_{i_k\cdots i_{\cal N}} = T_G\mathcal{D}_{i_k\cdots i_{\cal N}}.
\end{equation}
The number of observable events is given by the local rate density times the volume-time of the search. The combined number of NSNS and NSBH mergers is  
\begin{equation}
N_{i_1\cdots i_{\cal N}} = \left\{[\rho_G \mathcal{R}]T_{i_1\cdots i_{\cal N}}\mathcal{V}_{i_1\cdots i_{\cal N}}\right\}+\left\{[\rho_g(1-\mathcal{R})]T_{i_1\cdots i_{\cal N}}\mathcal{U}_{i_1\cdots i_{\cal N}}\right\},
\end{equation}
where $\rho_G$ is the combined local rate density of the NSNS and NSBH mergers, $\mathcal{R}$ is the fraction of NSNS mergers to the total number of NSNS and NSBH mergers, $\mathcal{V}_{i_1 \cdots i_{\cal N}}$ ($\mathcal{U}_{i_1 \cdots i_{\cal N}}$) is the second largest single-detector NSNS (NSBH) search volume during the time $T_{i_1\cdots i_{\cal N}}$ when two or more detectors are in observing mode (zero otherwise). By summing on all $i_k = (0,1)$ combinations, we find
\begin{eqnarray}
\label{eq:mainappendixB}
N_G &=& \sum_{i_1=0}^1\cdots \sum_{i_{\cal N}=0}^1 N_{i_1\cdots i_{\cal N}}\,, \nonumber \\
&=& \rho_G\sum_{i_1=0}^1\cdots  \sum_{i_{\cal N}=0}^1\left\{\mathcal{R}T_{i_1 \cdots i_{\cal N}}\mathcal{V}_{i_1 \cdots i_{\cal N}} + (1-\mathcal{R})T_{i_1 \cdots i_{\cal N}} \mathcal{U}_{i_1 \cdots i_{\cal N}}  \right\} \nonumber \\
&=& T_G\rho_G \sum_{i_1=0}^1\cdots  \sum_{i_{\cal N}=0}^1\left\{ \mathcal{D}_{i_1 \cdots i_{\cal N}}\left[\mathcal{R}\mathcal{V}_{i_1 \cdots i_{\cal N}} +  (1-\mathcal{R})\mathcal{U}_{i_1 \cdots i_{\cal N}} \right] \right\}\,.
\end{eqnarray}
As an example, we compute the number of NSNS mergers observable by the LIGO-Hanford and LIGO-Livingston network with duty cycles $D_1$, $D_2$ and search volumes $V_1\geq V_2$, respectively. Using Eq.\ (\ref{eq:mainappendixB}) we have
\begin{eqnarray}
N_{NSNS}& = &  T_G\rho_G\sum\limits_{i_1=0}^1\sum\limits_{i_2=0}^1\mathcal{D}_{i_1i_2}\mathcal{R}\mathcal{V}_{i_1i_2}\nonumber \\
& = & \rho_G\mathcal{R}T_G(\mathcal{D}_{00}\mathcal{V}_{00}+\mathcal{D}_{01}\mathcal{V}_{01}+\mathcal{D}_{10}\mathcal{V}_{10}+\mathcal{D}_{11}\mathcal{V}_{11})\nonumber \\
 &= &(\rho_G\mathcal{R})(T_GD_1D_2)V_2\,,
\end{eqnarray}
where $\mathcal{D}_{i_1i_2}$ and $\mathcal{V}_{i_1i_2}$ are given by 
\begin{eqnarray}
\mathcal{D}_{i_1i_2} &=& \prod_{k=1}^{2}\left[D_{k}^{i_k}(1-D_{k})^{1-i_k}\right] = D_1^{i_1}(1-D_1)^{1-i_1}D_2^{i_2}(1-D_2)^{1-i_2}\,,\\
  \mathcal{V}_{i_1i_2} &=&\left\{
  \begin{array}{@{}ll@{}}
     V_2 &\qquad \text{for}\qquad (i_1\,,i_2)=(1\,,1)\,, \\
    ~0 &\qquad \text{for} \qquad  (i_1\,,i_2)=(1\,,0)\,,(0\,,1)\,,(0\,,0)\,.\\
  \end{array}\right.
\end{eqnarray}

Similarly, the number $N_C$ of coincident events detectable by at least two GW detectors and an EM detector is obtained using ${\cal D}_{i_1\cdots i_{\cal N}}$ and $N(z)$ in Eq. (\ref{eq:cumnum}). If an EM detector with duty cycle factor $D_{\rm EM}$ and field of view $f_{\rm EM}$ is observing at the same time of $m$ detectors in the GW network, the number of observable SGRB events per year up to $z$ by the EM detector is given by $f_{\rm EM}N(z)/(fT_o)$ and the fraction of the observation time is $T_GD_{i_k\cdots i_{\cal N}}D_{\rm EM}$. The local rate densities of SGRBs which are produced by NSNS and NSBH mergers are given by
\begin{eqnarray}
    \rho_{S, NSNS} &=& \Gamma_{NSNS}\mathcal{R}\rho_G\,,\\
    \rho_{S, NSBH} &=& \Gamma_{NSBH}(1-\mathcal{R})\rho_G\,, 
\end{eqnarray}
where $\Gamma_{NSNS}$ and $\Gamma_{NSBH}$ are the fraction of NSNS and NSBH that produce SGRBs, respectively. The fraction of SGRBs which are produced by NSNS mergers is
\begin{equation}
    \mathcal{P} = \frac{\rho_{S, NSNS}}{\rho_S} = \frac{\Gamma_{NSNS}\mathcal{R}}{\Gamma_{NSNS}\mathcal{R}+\Gamma_{NSBH}(1-\mathcal{R})}
\end{equation}
where $\rho_S = \rho_{S, NSNS} + \rho_{S, NSBH}$ is the total SGRB local rate density. The number $N_C$ of coincident events detectable by at least two GW detectors and the EM detector is     
\begin{eqnarray}
N_C &=&  \sum\limits_{i_1=0}^1\cdots \sum\limits_{i_{\cal N}=0}^1 \left\{(T_G{\rm D}_{i_1\cdots i_{\cal N}} D_{\rm EM})\left[{\cal P}\frac{f_{\rm EM}N({\cal Z}_{i_1 \cdots i_{\cal N}})}{fT_o} + (1-{\cal P})\frac{f_{\rm EM}N({\cal Y}_{i_1 \cdots i_{\cal N}})}{fT_o}\right]     \right\}\nonumber \\ 
&=&\frac{T_G D_{{\rm EM}} f_{{\rm EM}}}{fT_o}\sum\limits_{i_1=0}^1\cdots\sum\limits_{i_{\cal N}=0}^1 \left\{\mathcal{D}_{i_1\cdots i_{\cal N}} \left[ \mathcal{P}N\left(\mathcal{Z}_{i_1\cdots i_{\cal N}}\right) + \left(1-\mathcal{P}\right)N\left(\mathcal{Y}_{i_1\cdots i_{\cal N}}\right)\right] \right\}\,.
\end{eqnarray}
The number of events detectable with only a single GW detector in coincidence with an EM detector can be obtained by setting $\mathcal{Z}_{i_1\dots i_{\mathcal{N}}}$ and $\mathcal{Y}_{i_1\dots i_{\mathcal{N}}}$ to the largest single-detector NSNS and NSBH search redshifts, respectively.

\end{document}